\documentclass[letterpaper]{article} 
\usepackage{aaai24}  
\usepackage{times}  
\usepackage{helvet}  
\usepackage{courier}  
\usepackage[hyphens]{url}  
\usepackage{graphicx} 
\usepackage{natbib}  
\usepackage{caption} 
\usepackage{algorithm}
\usepackage[noend]{algpseudocode}
\usepackage{amsmath}
\usepackage{amssymb}
\usepackage{booktabs}
\usepackage{xcolor}
\usepackage{pdfpages}
\usepackage{newfloat}
\usepackage{listings}
\DeclareCaptionStyle{ruled}{labelfont=normalfont,labelsep=colon,strut=off} 
\floatstyle{ruled}
\newfloat{listing}{tb}{lst}{}
\floatname{listing}{Listing}
\newcommand{\D}{\lq$\Delta$\rq}
\newcommand{\Dn}{\lq$\Delta_n$\rq}
\newcommand{\I}{\lq$I$\rq}
\newcommand{\ID}{\lq$I \cup \Delta$\rq}
\newcommand{\TD}{$\Delta^T$}
\newcommand{\TI}{$I^T$}

\newcommand{\scheduleF}{\textit{schedule}}
\newcommand{\insertF}{\textit{derive}}
\newcommand{\mergeF}{\textit{merge}}

\newcommand{\PLH}{{\mkern-2mu\times\mkern-2mu}}

\newtheorem{theorem}{Theorem}

\newcommand{\mysubsection}[1]{\smallskip\noindent\textbf{#1}:}
\newcommand{\inappendix}{appendix}

\title{Optimised Storage for Datalog Reasoning}
\author {
    Xinyue Zhang\textsuperscript{\rm 1},
    Pan Hu\textsuperscript{\rm 2},
    Yavor Nenov\textsuperscript{\rm 3},
    Ian Horrocks\textsuperscript{\rm 1}
}
\affiliations {
    \textsuperscript{\rm 1}Department of Computer Science, University of Oxford, Oxford, UK\\
    \textsuperscript{\rm 2}School of Electrical Information and Electronic Engineering, Shanghai Jiao Tong University, China\\
    \textsuperscript{\rm 3}Oxford Semantic Techonologies, Oxford, UK \\
    \{xinyue.zhang, ian.horrocks\}@cs.ox.ac.uk, pan.hu@sjtu.edu.cn, yavor.nenov@oxfordsemantic.tech
}

\usepackage{bibentry}

\begin{document}

\maketitle

\begin{abstract}
Materialisation facilitates Datalog reasoning by precomputing all consequences of the facts and the rules so that queries can be directly answered over the materialised facts. However, storing all materialised facts may be infeasible in practice, especially when the rules are complex and the given set of facts is large. We observe that for certain combinations of rules, there exist data structures that compactly represent the reasoning result and can be efficiently queried when necessary. In this paper, we present a general framework that allows for the integration of such optimised storage schemes with standard materialisation algorithms. Moreover, we devise optimised storage schemes targeting at transitive rules and union rules, two types of (combination of) rules that commonly occur in practice. Our experimental evaluation shows that our approach significantly improves memory consumption, sometimes by orders of magnitude, while remaining competitive in terms of query answering time. 
\end{abstract}

\section{Introduction}
Datalog~\cite{abiteboul1995foundations} can describe a domain of interest as a set of ``if-then" rules and new facts in this domain can be derived by applying the rules to a set of explicitly given facts until a fixpoint is reached.
With the ability to express recursive dependencies, such as transitive closure and graph reachability, Datalog is widely used in different communities. 
In the Semantic Web community, Datalog is used to capture OWL 2 RL ontologies~\cite{motik2009owl} possibly extended with SWRL rules~\cite{horrocks2004swrl} and can thus be used to answer queries over ontology-enriched data.
There are an increasing number of academic and commercial systems that have implemented Datalog, such as LogicBlox~\cite{aref2015design}, VLog~\cite{carral2019vlog}, RDFox~\cite{nenov2015rdfox}, Vadalog~\citep{bellomarini2018vadalog}, GraphDB\footnote{\url{https://graphdb.ontotext.com/}}, and Oracle's database~\cite{wu2008implementing}.

Given a set of explicitly given facts and a set of Datalog rules, a prominent computational task for a Datalog system is to answer queries over both facts and rules.
One typical approach is to pre-compute all derivable facts from the rules and original facts.
This process of computing all consequences is known as \textit{materialisation}, the same for the resulting set of facts.
Materialisation ensures efficient query evaluation, as queries can be directly evaluated over the materialised facts without considering the rules further. 
Therefore, materialisation is commonly used in Datalog systems. For example, systems like RDFox, Vadalog, LogicBlox, and VLog all adopt this approach.  
However, materialisation has downsides for large datasets. Computing the materialisation can be computationally expensive, especially with rules like transitive closure that derive many inferred facts. Storing all the materialised facts also increases storage requirements. 
Additionally, if the original facts change, materialisation needs to be incrementally updated, rather than fully recomputed from scratch each time. This incremental maintenance is crucial for efficiency when facts are updated frequently. 
In essence, materialisation trades increased preprocessing time and storage for improved query performance by avoiding extensive rule evaluation during query processing. 
The costs in time and space to materialise can become prohibitive for very large datasets and rule sets. 

The computation and maintenance of materialisation have been well investigated. The standard \textit{semina\"ive} algorithm~\cite{abiteboul1995foundations} efficiently computes the materialisation by avoiding repetitions during rule applications.
This algorithm can also incrementally maintain the materialisation for fact additions. 
More general (incremental) maintenance algorithms like the \textit{counting} algorithm~\cite{gupta1993maintaining}, \textit{Delete/Rederive} algorithm~\cite{staudt1995incremental}, and \textit{Backward/Forward} (B/F) algorithm~\cite{motik2019maintenance} can maintain materialisation for both additions and deletions and are applicable beyond initial computation. 
Additionally, specialised algorithms optimised for particular rule patterns, like transitive closure~\cite{subercaze2016inferray}, have been developed, and a modular framework proposed by~\citet{hu2022modular} combines standard approaches for normal rules with tailored approaches for certain types of rules to further improve materialisation efficiency. 

While extensive research has been conducted on the efficient computation and maintenance of Datalog materialisation, 
%
%
optimised storage of relations that takes into account properties implied by the program 
has so far been limited to the handling of equality relations \cite{motik2015handling} and the exploitation of columnar storage \cite{carral2019vlog}.
However, traditional materialisation methods become impracticable due to oversized fact repositories and rule sets that significantly expand data volume.
For example, materialising the transitive closure of just the \textit{broader} relation in DBpedia~\cite{DBLP:journals/semweb/Lehmann2015DBpedia} results in 8.5 billion facts, which would require an estimated 510 GB of memory to store.
The failure of materialisation makes further query answering unachievable. 
In this work, we investigated tailored data structures to minimise memory utilisation during materialisation, focusing on transitive closure and union rules.
We proposed non-trivial approaches for efficiently handling incremental additions of specialised data structures, an unavoidable and essential step in Datalog Reasoning.
Additionally, we proposed a general multi-scheme framework that separates storage from reasoning processes, allowing for various storage optimisations.
Overall, this research aims to provide a novel way to process large fact sets and Datalog rules. It lays the groundwork for using materialisation to store and query large knowledge graphs efficiently.

This paper is organised as follows. First, we introduce some preliminary concepts and background. We then present our general framework for reasoning over customised data structures. Next, we detail specific optimised data structures for materialisation, including methods for initial construction and incremental maintenance under fact additions.
Finally, we empirically evaluate our techniques, demonstrating improved performance and reduced memory usage compared to standard materialisation approaches, including cases where traditional materialisation fails.
Additionally, we empirically evaluate query performance over our optimised data schemes. 
For small queries, response times using our tailored structures are comparable to plain fact storage, demonstrating efficient access. 
The evaluations highlight the benefits of tailored data structures and reasoning algorithms in enabling efficient large-scale materialisation. The datasets and test systems are available online\footnote{https://xinyuezhang.xyz/TCReasoning/}. 

\section{Preliminaries}
\mysubsection{Datalog}
A \textit{term} is a variable or a constant. An \textit{atom} has the form $P(t_1, \dots, t_k)$, where $P$ is a predicate with arity $k$ and each $t_i$ is a term. A \emph{fact} is a variable-free atom, and a \textit{dataset} is a finite set of facts.
%
A \textit{rule} is an expression of the form: 
    $B_0 \wedge \dots \wedge B_n \rightarrow H$,
where $n \geq 0$ and $B_i$, $0 \leq i \leq n$, and $H$ are atoms. For $r$ a rule, $\mathsf{h}(r) = H$ is its \textit{head}, and $\mathsf{b}(r) = \{B_0, \dots, B_n \}$ is the set of \textit{body atoms}. 
A rule is \textit{safe}, if each variable in its head atom also occurs in some of its body atoms.
A \textit{program} is a finite set of safe rules.

A \emph{substitution} is a finite mapping of variables to constants. Let $\alpha$ be a term, an atom, a rule, or a set thereof. The \emph{application} of a substitution $\sigma$ to $\alpha$, denoted as $\alpha \sigma$, is the result of replacing each occurrence of a variable $x$ in $\alpha$ with $\sigma(x)$, if $x$ is in the domain of $\sigma$.
For a rule $r$ and a substitution $\sigma$, if $\sigma$ maps all the variables occurring in $r$ to constants, then $r\sigma$ is an \textit{instance} of $r$.

For a rule $r$ and a dataset $I$, 
$r[I] = \{\mathsf{h}(r\sigma)\ |\ \mathsf{b}(r\sigma) \subseteq I \}$ is the set of facts obtained by applying $r$ to $I$. Given a program $\Pi$, $\Pi[I] = \bigcup_{r\in \Pi} \{ r[I] \}$ is the result of applying every rule $r$ in a program $\Pi$ to $I$.
%
%
The \textit{materialisation} $I_\infty$ of $\Pi$ w.r.t. a dataset $E$ is defined as $I_\infty = \bigcup_{i\geq 0} I_i$ in which $I_0 = E$, $I_{i}  = I_{i-1} \cup \Pi[I_{i-1}] \text{ for } i > 0$. 
Similarly, let $\Pi^i[I_0] = I_i$ be the facts inferred by applying rules in $\Pi$ to initial facts $I_0$ and recursively to previous inferred facts, for $i$ iterations.

\mysubsection{Semina\"ive algorithm}
The semina\"ive algorithm shown in Algorithm~\ref{alg:semi} performs Datalog materialisation without repetitions of rule instances. 
The set $E$ and $I$ are initialised as empty. In the initial materialisation, the dataset is given to $E^+$.
The $\Delta$ is first initialised as $E^+$. In each round of rule application, new facts $\Delta$ is used by the operator $\Pi[I, \Delta]  = \bigcup_{r\in \Pi} \{ r[I, \Delta ] \}$, in which $\Delta \nsubseteq I$, and $r[I,\Delta ]$ is defined as follows:
\begin{equation}
    r[I,\Delta ] = \{ \mathsf{h}(r\sigma) \ | \  \mathsf{b}(r\sigma)  \subseteq I \cup \Delta \ \text{,}\  \mathsf{b}(r\sigma) \cap \Delta \neq \emptyset\}, \label{equ:rSemiNaive} 
\end{equation}
in which $\sigma$ is a substitution mapping variables in $r$ to constants.
The definition of $\Pi[I, \Delta]$ ensures the algorithm only considers rule instances that are not considered in previous rounds. 
Then in line~\ref{alg:semi:merge}, $\Delta$ is merged to $I$ and new derivations $A$ found in the current round are assigned to $\Delta$ to be used in the next round.
The incremental addition is processed similarly by initialising $E^+$ as the facts to be inserted. 
Similar to the semina\"ive algorithm, we also identify facts in the domain \I\ and \D\ when processing the materialisation. 
%


\mysubsection{Modular reasoning}
\citet{hu2022modular} present a modular version of the semina\"ive algorithm, which integrates standard rule application with the optimised evaluation of certain rules (e.g. transitive closure and chain rules). 
A Datalog evaluation is split into modules each of which manages a subset of the original program. 
The modular semina\"ive algorithm is then obtained by replacing line~\ref{alg:semi:ruleApply} in Algorithm~\ref{alg:semi} with $A = A \cup (\Pi_T^+[I, \Delta] \setminus \Delta )$ for every module $T$, where $\Pi_T[I, \Delta] \setminus I \subseteq \Pi_T^+[I, \Delta] \subseteq \Pi_T^\infty[I \cup \Delta] \setminus I$.

\begin{algorithm}[t]
\caption{Seminaive($\Pi, I, E, E^+$)}\label{alg:semi}
\begin{algorithmic}[1]
\State \textbf{Result:} update $I$ from $\Pi^\infty[E]$ to $\Pi^\infty[E\cup E^+]$
\State $\Delta := E^+ \setminus E $ \label{alg:semi:initD}
\While{$\Delta \neq \emptyset$}                         \label{alg:semi:whileBeg}                         
    \State $A := \Pi[I, \Delta] \setminus (I \cup \Delta)$  \label{alg:semi:ruleApply}
    \State $I:=I\cup \Delta$ \label{alg:semi:merge}
    \State $\Delta := A $ \label{alg:semi:AtoD}
\label{alg:semi:whileEnd}
\EndWhile
\end{algorithmic}
\end{algorithm}


\section{Motivation}

In this section, we illustrate the benefits of using a specialised storage scheme for Datalog reasoning. Let $\Pi$ be the program containing the rule $R(x,y), R(y,z) \rightarrow R(x,z)$ that declares a binary relation $R$ as a transitive relation. Let $E^+$ be the set of facts ${\{ R(a_{i+1}, a_i) \ |\ 1 \leq i \leq n - 1 \}}$. The materialisation obtained by applying $\Pi$ to $E^+$ is ${I = \{ R(a_i, a_j)\ |\ 1 \leq j < i \leq n\}}$. Each fact ${R(a_i, a_j)}$ in ${I \backslash E^+}$ can be derived $i-j-1$ times by rule instance $ R(a_i, a_k), R(a_k, a_j) \rightarrow R(a_i, a_j)$ for $k$ with $j < k < i$.
%
The semina\"ive algorithm considers each distinctive and applicable rule instance once, so the materialisation requires $O(n^3)$ time. \citet{hu2022modular} proposed an optimisation that requires one of the body atoms to be matched in the explicitly given facts, thus avoiding considering all applicable rule instances and lowering the running time to $O(n^2)$ on this input. In both cases, storing the materialised result clearly requires $O(n^2)$ space. 

\begin{figure}
    \centering
    \includegraphics[width=0.8\columnwidth]{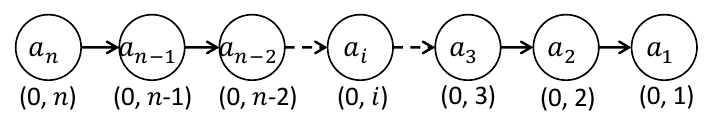}
    \caption{The chain with associated intervals.}
    \label{fig:chain}
\end{figure}

We next outline an approach that requires significantly less space and is at the same time efficient to compute. Our approach builds upon the transitive closure compression technique by \citet{agrawal1989efficient}, which makes use of interval trees to allow for compact storage and efficient access of transitive relations. 
Treating each constant appearing in $E^+$ as a node and each fact as a directed edge, the idea is to assign each node $v$ an index and an interval such that the interval covers exactly the indexes of the nodes that are reachable from $v$. 
An example interval tree representing the transitive relation over $E^+$ is depicted in Figure~\ref{fig:chain}.
Then, $I_{a_i}$, facts in the closure with $a_i$ in the first position can be retrieved using indexes $id$ and intervals $In$: 
\begin{align}
    I_{a_i} & = \{ R(a_i, a_j) \ |\ a_j.id \in a_i.In \} . \label{exp:example:Iai} 
\end{align}
The full materialisation is in essence $\{ I_{a_i} \ |\ 1 \leq i \leq n \}$. Answering point queries such as whether $R(a_i, a_j)$ holds could also be efficiently implemented: it suffices to check whether $a_j$'s index is covered by $a_i$'s interval, an operation that requires $O(1)$ time.

The above data structure can be computed by performing a post-order traversal starting from the root $a_n$. When a node is visited, its index is assigned by increasing the index of the previous node by one, and its interval is computed using the interval of its child. This requires only $O(n)$ time, as opposed to $O(n^2)$ and $O(n^3)$ required by existing approaches. In terms of space usage, 
in our particular example, the data structure requires $O(n)$ space, as opposed to $O(n^2)$ required by a full materialisation. For an arbitrary graph in general, 
the corresponding indexes and intervals can be constructed in $O(|V| + |E|)$ time, where $|V|$ and $|E|$ are the numbers of vertices and edges of the graph, respectively, and the worst-case space complexity is $O(|V|^2)$. Note that space consumption can sometimes be reduced by choosing the optimum tree cover of a graph~\cite{agrawal1989efficient}, a technique that proves to be useful in our evaluation.

The above example shows that a customised storage scheme saves time and space during construction, and also efficiently supports query answering. 
However, in typical application scenarios, a Datalog program $\Pi$ usually contains multiple different rules, and different optimisations may apply. 
How to combine different storage schemes and enable their integration with standard reasoning algorithms remains a challenge. In our example, $\Pi$ may include other rules that also derive $R$ facts, and it is essential that the interaction between such rules and the transitive rule is properly handled. 
Moreover, to enable semina\"ive evaluation, the optimised storage schemes should provide efficient access to different portions of the derived facts (i.e., \I, \D, \ID), which will involve nontrivial adaptation of existing approaches. 

We address the above issues by first introducing a general framework which involves the specification of several interfaces that each storage scheme must implement. We then present details of two useful storage schemes, focusing on the implementation of the relevant interfaces.

\section{Multi-Scheme Framework}

\begin{algorithm}[t]
\caption{Multi-SchemeReasoning($\Pi, E, E^+$)}
\label{alg:reason}
\begin{algorithmic}[1]
    \State \textbf{Result}: update $I$ from $\Pi^\infty[E]$ to $\Pi^\infty[E\cup E^+]$ 
    \State \scheduleF$(E^+ \backslash E)$  \Comment{populate $\Delta_n^T$ and $C_T$} \label{alg:reason:initSchedule}
    \Loop  
        \State $\Delta =$ \insertF() \label{alg:reason:insert} \Comment{apply $\Pi_T$ and update $\Delta^T$}
        \If{$\Delta = \emptyset$}{ break}
        \EndIf
        \For{every scheme $T$} 
            \State \scheduleF(\TD) \label{alg:reason:schedule}  \Comment{populate $\Delta_n^T$ and $C_T$}
            \State $T$.\mergeF() \label{alg:reason:merge} \Comment{merge \TD\ to \TI, empty \TD}
        \EndFor
    \EndLoop
\end{algorithmic}
\end{algorithm}

Our framework incorporates multiple storage schemes that are responsible for managing disjoint sets of facts. In particular, each storage scheme $T$ deals with facts corresponding to predicates appearing in  $P_T$, and it is associated with rules $\Pi_T \subseteq \Pi$ that use predicates in $P_T$ in the head. Additionally, to facilitate representation, we use another set of predicates $SP_T$ to denote the predicates used in the body of rules in $\Pi_T$. Moreover,
an internal data structure $D_T$ maintains facts in $T$ and a fact cache $C_T$ is used to temporarily store the input facts.
Finally, we denote by \TI the facts in the domain \I\ managed by scheme $T$. Similarly, we denote by \TD the facts in the domain \D\ managed by scheme $T$.
To work with different schemes correctly during the materialisation, each scheme should implement the following functions.
\begin{enumerate}
    \item The \textit{schedule} function identifies facts with predicate in $SP_T$, and stores them in $C_T$ so that these facts can be used by $\Pi_T$ to derive new facts. An input fact $t$ with the predicate in $P_T$ is added to a set $\Delta_n^T$ if $t \notin $ \TD $\cup$ \TI. This function does not change \TD or \TI for a scheme $T$; it only schedules facts for later computation of \TD.
    
    \item The \insertF\ function applies rules in $\Pi_T$ and incorporates new facts in the data structure. The function does not modify \TI\ but updates \TD\ as follows.
    \begin{align}
        \text{\TD}  &= \Delta_n^T \cup \Pi_T^+[I, C_T] . \label{exp:TD:gt} 
    \end{align}
    \item The \mergeF\ function updates \TI to \TI$\cup$\TD, empties \TD.
\end{enumerate}
%
The global \scheduleF\ function invokes \scheduleF\ functions of every scheme. 
The global \insertF\ function calls \insertF\ functions of every scheme, and returns all facts in domain \D.
%
The reasoning algorithm incorporating multiple storage schemes is shown in algorithm~\ref{alg:reason}. It exploits principles similar to the modular materialisation approach.
The main difference is that our approach additionally manages the (possibly compact) representation of derivations for different parts of the program.
In line~\ref{alg:reason:initSchedule}, relevant facts are identified for each scheme. 
In line~\ref{alg:reason:insert}, rules in $\Pi_T$ are applied in each scheme, and the data structure $D_T$ is updated to incorporate facts in $\Delta_n^T$ and the newly derived consequences into $\Delta^T$.
Then in line~\ref{alg:reason:schedule}, $\Delta^T$ are scheduled for insertion into different schemes before being merged to \TI\ in line~\ref{alg:reason:merge}.
In contrast to the modular materialisation approach in which a module $T$ computes only $\Pi_T^+[I, \Delta]$, our approach additionally considers $\Delta_n^T$ as part of $\Delta^T$ in~\eqref{exp:TD:gt}. 
This is to make our framework general enough to capture storage schemes that do not explicitly store facts and thus cannot easily distinguish between their input and consequences. As we shall see, our storage scheme for transitive relations benefits from this generalisation. The following theorem states that algorithm~\ref{alg:reason} is correct, and its proof is provided in the \inappendix.
\begin{theorem}
    A fact can be derived and represented in relevant schemes by the multi-scheme algorithm if and only if it can be derived by the modular semina\"ive algorithm.
\end{theorem}

\mysubsection{Plain Table}
In practice, predicates and rules that are not handled by customised storage schemes are managed by a plain table. The plain table, as the name suggests, stores facts faithfully without any optimisations.
The internal data structure $D_T$ can be implemented, for example, as a fact list $L_T$, in which each fact is assigned a label, either \D\ or \I. 
Then \TI\ and \TD\ are defined intuitively as facts with the corresponding label. 
The \insertF\ function adds facts in $\Delta_n^T$ to $L_T$ and marks them as \D. Also, derivations in $\Pi_T[I, C_T]$ are added to the list $L_T$ with the label \D\ if they are not in the list.
The \mergeF\ function is realised by simply changing the label of facts. 
It is easy to verify that the plain table satisfies the requirement of a scheme.

\section{TC Storage Scheme}
This section presents a specialised transitive closure (TC) scheme capable of efficiently handling transitively closed relations. The implementation of the scheme's functions is based on nontrivial adaptations of the interval-based approach by~\citet{agrawal1989efficient}, which treats TC computation as solving reachability problems over a graph. More specifically, each node is assigned an interval that compactly represents the (indexes of) nodes it can reach. The original approach does not accommodate access to facts in various domains (i.e., \I, \D, \ID). Furthermore, their discussion of incremental updates does not encompass all cases. Our extension of the technique enables multi-scheme reasoning by supporting access to different domains and providing more comprehensive incremental update procedures. 

\mysubsection{Interval-based Approach}
For a set of input facts represented by a graph $G$, the approach computes a tree cover of the graph. Each node of the graph is numbered based on the post-order traversal of the tree. 
Then, an initial interval is assigned to each node with its post-order index being the upper bound, and the smallest lower-end number among its descendants' intervals being the lower bound. 
For leaves, the lower bound is its index.
The initial intervals capture the reachability of the tree. Then, for every edge $(i,j)\in G$ that is not in the tree cover, the interval of $j$ is added to $i$ and all its ancestors. 
The final intervals capture all reachable pairs in $G$. 
Just as expression~(\ref{exp:example:Iai}), for each node in $G$, facts in the closure with this node as the first constant can be accessed using the computed intervals and indexes.

\mysubsection{Settings}
For the remainder of this section, we assume there is a rule $r \in \Pi$ that axiomatises a relation $R$ as transitive; the predicate set $P_T$ contains $R$ (and so $r$ is in $\Pi_T$). Additionally, 
for the ease of presentation, we assume that $\Pi_T$ contains only $r$. In reality, $T$ could also handle other rules that derive $R$ facts: these rules are applied using a standard algorithm, and the output is stored and processed by $T$. 

\subsection{Incremental Update and Fact Access}
We now discuss how customised storage schemes deal with incremental insertion, which is crucial for integrating with standard reasoning algorithms. Notice that the original approach by~\citet{agrawal1989efficient} already considered incremental insertion. However, their discussion does not cover all possible insertion cases and the distinction between facts in domains \I\ and \D\ is not allowed, which is a key requirement in the Datalog reasoning setting. 

\mysubsection{Naive Approach}
One seemingly straightforward solution to supporting the distinction between \I\ and \D\ facts is to have two sets of intervals for each node $s$: $s.In$ and $s.D$ contain indexes that $s$ can reach before the insertion and that $s$ can additionally reach after the insertion, respectively.
Let $I_s$ and $\Delta_s$ be the facts with $s$ as the first constant in \I\ and \D, respectively, then \TD\ and \TI\ can be defined as follows:
\begin{align}
     I_s & = \{ R(s, x) \ |\ x.id \in s.In \},  \\
    \Delta_s & = \{ R(s, x) \ |\ x.id \in s.D \}, \\
    \text{\TD} & = \{ \Delta_s \ |\ s \in G \}, \text{\TI} = \{ I_s \ |\ s \in G \} . \label{exp:method1:TITD}
\end{align}
The \mergeF\ function merges \D\ to \I\ by simply adding $s.D$ to $s.In$ and emptying $s.D$ afterwards. 

\begin{figure}
    \centering
    \includegraphics[width=0.6\columnwidth]{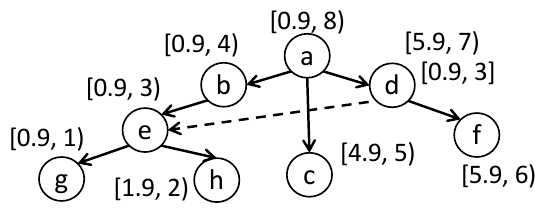}
    \caption{One example interval tree. The dashed edge is an edge in the graph but not covered by the tree cover.}
    \label{fig:intervalTree}
\end{figure}

\mysubsection{The Problem of Fresh Nodes}
We use an example to show why the naive approach is insufficient in the presence of fresh nodes. Assume that we insert a fresh node $k$ and a new edge $(d, k)$ to the graph shown in Figure~\ref{fig:intervalTree}. The tree cover is supposed to cover all the nodes, so $(d, k)$ is added to the tree cover. 
Instead of re-assigning the post-order index based on the updated tree, we assign a new post-order number to $k$ by finding a number $i$ so that $i$ is included in the initial interval of $d$ and $i$ is not occupied by any existing nodes, as suggested by~\citet{agrawal1989efficient}. 
After insertion, the intervals would still be valid if we did not intend to distinguish between `\I' and `\D'. However, following the naive approach we have $k.id \in d.In$, and so $R(d,k) \in $ \TI, which is not as expected since $(d, k)$ is newly inserted. 

We propose to have another set of intervals $s.N$ to memorise the reachable nodes that are freshly introduced, where $s.N \subseteq s.In \cup s.D$. So new nodes in $s.In$ can be identified and skipped when accessing \TI, and they can be returned in addition to nodes in $s.D$ when accessing \TD. Formally:
\begin{align}
     I_s & = \{ R(s, x) \ |\ x.id \in s.In \backslash s.N \},  \\
    \Delta_s & = \{ R(s, x) \ |\ x.id \in s.D \cup s.N  \},
\end{align}
in which we treat a set of intervals as the set of numbers it covers, and set operators such as intersection ($\cap$), union ($\cup$ or +) and minus ($\backslash$) naturally apply. 
The \TD\ and \TI\ are defined in the same way as in expression~\eqref{exp:method1:TITD}. 
Notice that to allow for the insertion of edges involving fresh nodes, when creating the initial intervals, we should allow for gaps. This could be achieved, for example, through special treatment for the leaf nodes: the lower end of the interval is set to be the current node index minus a small margin (e.g., 0.1), as illustrated in Figure~\ref{fig:intervalTree}.

\mysubsection{Graph with Cycles}
While the previous discussion assumes the graph constructed from the input facts is acyclic in the initial construction of the data structure, the same technique can also be applied to a cyclic graph by collapsing each strongly connected component to a node. Let $G_0$ and $G$ be the graph before and after condensation, respectively, and let $M$ be a mapping that maps each SCC in $G$ to its corresponding nodes in $G_0$. Then, the indexes and intervals are assigned to SCCs of the graph after condensation in the same way as described above. For each $s \in G$, sets $I_s$ and $\Delta_s$ denoting the corresponding sets of facts from $G_0$ in domains \I\ and \D\ , respectively, can be obtained as follows:
\begin{align}
     I_s & = \{ M(s)\PLH M(x) |\ x.id \in s.In \backslash s.N \}, \label{exp:stable:In} \\
    \Delta_s & = \{ M(s)\PLH M(x) |\ x.id \in s.D \cup s.N \}, \label{exp:stable:Dn} 
\end{align}
in which $M(s) \PLH M(x)$ computes the cross-product between the two sets of constants $M(s)$ and $M(x)$; for brevity, the predicate $R$ is omitted. 
Finally, \TD\ and \TI\ are the same as in expression~\eqref{exp:method1:TITD}. 

\mysubsection{The Merging of Components}
A tricky case not discussed in the original approach is that fact additions could possibly lead to the merging of existing SCCs. For example, if an edge $(h, d)$ is introduced, then the SCC $e, h$ and $d$ need to be merged as a new SCC. 
The graph can be updated by choosing $e$ as a representative node and deleting nodes $h$ and $d$. 
The children of $h$ and $d$ will be inherited to $e$. 
However, it is not straightforward how to maintain the intervals and access facts in \TI\ and \TD\ correctly. 
For SCCs that are not merged during fact additions, expressions~(\ref{exp:stable:In}) and~(\ref{exp:stable:Dn}) are still valid to use. 
We distinguish between different SCCs by their status $St$: the status of an SCC that is not merged is \textit{stable}; the status of an SCC that is merged and selected to be representative is \textit{new}; the status of an SCC that is merged to a representative SCC is \textit{dropped}. 
For a \textit{new} SCC $e$, we use $e.D$ to store the interval that includes the indexes of nodes that $e$ can reach after merging regardless of the domain, and $e.N$ includes newly introduced nodes in $e.D$, so $e.N \subseteq e.D$. 
Intuitively, the node $e$ after merging is able to reach all the nodes in this newly merged component, as well as their descendants, so $e.D = \bigcup_{s \in F(e)}\{ s.In \cup s.D \cup [s.id] \}$, in which $F$ is the map from the SCCs in the new graph to the original ones, and $[s.id]$ is a singleton interval that only includes $s.id$. 
In this example, $F(e) = \{ e, h, d \}$.
Similarly, $e.N$ also takes the union of $s.N$ for $s \in F(e)$, i.e., $e.N =  \bigcup_{s \in F(e)} \{ s.N \} $. 
Let $L$ be a list of SCCs ordered by their post-order indexes. Intervals and other associated information of SCCs are stored in $L$. 
Instead of deleting \textit{dropped} SCCs in $L$ right away, we keep the original $In$, $id$, and $M$ map of each node $s \in F(e)$ in $L$.
In this way, $\Delta_s$ and $I_s$ can be recomputed as follows:
\begin{align}
    I_s &= \{M(s) \PLH M(x)|\ x.id \in s.In \backslash e.N \}, \\
    \Delta_s &= \{ M(s) \PLH M(x)|\ x.id \in (e.D\backslash s.In)\text{+}(e.N \cap s.In) \}, \nonumber 
\end{align}
in which $e$ is the representative node of merged components so that $s \in F(e)$. 
In the \mergeF\ function of the TC storage scheme, \textit{dropped} SCCs in $L$ are deleted, $F$ map is emptied, and $M$ map of \textit{new} nodes is updated to the union of members of original SCCs. 
Moreover, the status of \textit{new} nodes is changed to \textit{stable}, and $D$ interval is merged to $In$, $N$ and $D$ intervals are emptied. 

The designs discussed above make the TC storage scheme suitable to use in the multi-scheme reasoning algorithm. For brevity and ease of understanding we have only highlighted the key aspects of the approach. Readers interested in an exhaustive (and lengthy) presentation of the algorithmic details of the storage scheme are invited to consult the \inappendix.

\section{Union Storage Scheme}

The union storage scheme is motivated by rules with the form: $A(x,y) \rightarrow U(x,y), B(x,y) \rightarrow U(x,y) .$
The facts with predicate $U$ can be derived by `copying' the instantiations from $A$ and $B$ facts. Therefore, instead of deriving and storing all the consequences of the above rules, we can have a `virtual' storage for the facts with predicate $U$. 
Assume we have a union table $T$ for $U$; $\Pi_T$ contains all the rules $r \in \Pi$ such that $r$ is of the form $p(x,y) \rightarrow U(x,y)$; for brevity we again assume that there is no other rule in $\Pi$ that derives $U$ facts; the set $SP_T$ contains predicates used in the body of rules in $\Pi_T$.
For the above example, $SP_T = \{A, B\}$.
The internal data structure in the union table is a fact list $L_T$: if a fact with $U$ is explicitly defined and cannot recover from $\Pi_T$, then this fact will be stored in $L_T$.

For a predicate $p$, $I_p$ and $\Delta_p$ denote the facts with predicate $p$ in corresponding domains.
The function responsible for computing \TI, as well as the implementation of the interfaces required by our framework, is described in algorithm~\ref{alg:union}.
The $U$ facts are the `union' of facts with predicates in $P_T$ and $SP_T$, so the \TI\ set first collects explicit facts in $L_T$ with label \I. Then, the remaining facts are translated from the supporting facts belonging to the same domain. Note that the function call $U(\vec x).sub(t)$ obtains the instantiation from $t$ and uses it to instantiate $U(\vec x)$.
In the \scheduleF\ function, only facts with the predicate in $P_T$ or $SP_T$ are relevant for processing. If the fact passes the relevance check but cannot be recovered from $I_p$ (line~\ref{alg:union:scheduleCond}), then the corresponding $U$ fact must be new and should be included in domain \D\ in the derivation stage. To prepare for such derivation, there are two distinct cases: if $t$ has predicate $U$, then $t$ is added to the list $L_T$ with label \Dn\ (line~\ref{alg:union:schd:pt}); if $t$ has predicate appearing in $SP_T$, then a $U$ fact instantiated by $t$ is added to $C_T$ for later use (line~\ref{alg:union:schd:spt}).
The \insertF\ function changes facts with the label \Dn\ to \D. 
The set \TD\ includes the \D\ facts in $L_T$, and translated facts in $C_T$.
The use of $C_T$ in \insertF\ and \scheduleF\ is only for the sake of better presentation. In reality, the fact translation is done on the fly and thus does not incur significant memory overhead.
The \mergeF\ function is realised by changing facts that are explicitly stored in $L_T$ with the label \D\ to \I.
It can be verified that the above implementation satisfies the definition of a scheme, and the proof is provided in the \inappendix.

\begin{algorithm}[t]
\caption{Functions of Union Table}
\label{alg:union}
\begin{algorithmic}[1]
    \State $T$: a union table; \quad $t$: a fact;  \quad $L_T$: the fact list in $T$.
    \State $D_T:$ implemented as $L_T$. \quad Assume $P_T = \{U\}$.
    \State $\Pi_T = \{ r\ |\ r = p(\vec x) \rightarrow U(\vec x) \in \Pi \}$. \label{alg:union:pit}
    
    \Procedure{compute \TI}{}
        \State \TI $ := \{ t\in L_T\ |\ \text{the label of } t =$ \I $\}$
        \For{$p \in SP_T$, \textbf{for} $t \in I_p$}  \State \TI $ := \text{\TI} \cup \{ U(\vec x).sub(t)\}$
        \EndFor
    \EndProcedure


    \Procedure{$T.$\scheduleF}{$t$}
        \If{$t.p \notin P_T \cup SP_T$}{ return}
        \EndIf 
        \If{$p(\vec x).sub(t)$ $\notin I_p$ for any $p \in SP_T\cup P_T$} \label{alg:union:scheduleCond}
            \If{$t.p \in P_T$}{ add $t$ to $L_T$ as \Dn} \label{alg:union:schd:pt}
            \EndIf
            \If{$t.p \in SP_T$}{ $C_T := C_T \cup \{ U(\vec x).sub(t) \}$} \label{alg:union:schd:spt}
            \EndIf
        \EndIf
    \EndProcedure


    \Procedure{$T.$\insertF()}{}
        \State mark facts in $L_T$ with \Dn\ as \D
        \State \TD $ := \{ t\in L_T\ |\ \text{the label of } t =$ \D $\}$
        \State \TD $:= \Delta^T \cup C_T$, \quad $C_T = \emptyset$
    \EndProcedure
    
\end{algorithmic}
\end{algorithm}

\section{Evaluations}

\begin{table*}[t]
\centering
\resizebox{\textwidth}{!}{%
\begin{tabular}{l|rrr|rrr|rrr|rrr}
\toprule
 & \multicolumn{3}{c|}{\textbf{159,561 $\triangleright$ 29,086,642}} & \multicolumn{3}{c|}{\textbf{29,086,642 $\triangleright$ 29,807,285}} & \multicolumn{3}{c|}{\textbf{29,807,285 $\triangleright$ 30,717,549}} & \multicolumn{3}{c}{\textbf{30,717,549 $\triangleright$ 53,928,267}} \\ 
 \midrule
 & time & \multicolumn{1}{c}{peak} & \multicolumn{1}{c|}{static} & time & \multicolumn{1}{c}{peak} & \multicolumn{1}{c|}{static} &  time & \multicolumn{1}{c}{peak} & \multicolumn{1}{c|}{static} &  time & \multicolumn{1}{c}{peak} & \multicolumn{1}{c}{static} \\ \midrule 
Standrad & \multicolumn{1}{r}{2,845.08} & 1,480.58 & 1,251.77 & \multicolumn{1}{r}{96.02} & 1,483.22 & 1,275.57 & \multicolumn{1}{r}{92.07} & 1,483.22 & 1,305.44 & \multicolumn{1}{r}{6,352.03} & 2,750.76 & 2,340.10 \\
TCModule & 28.24 & 1,554.44 & 1,320.64 & 15.50 & 1,564.41 & 1,358.62 & 2.78 & 1,564.41 & 1,374.29 & 39.42 & 2,845.42 & 2,414.81 \\
TCScheme & \multicolumn{1}{r}{8.44} & 249.64 & 248.01 & \multicolumn{1}{r}{0.84} & 249.74 & 248.01 & \multicolumn{1}{r}{0.91} & 249.77 & 248.01 & \multicolumn{1}{r}{6.06} & 284.93 & 278.26 \\ 
\midrule
 & \multicolumn{3}{c|}{\textbf{1,426,588 $\triangleright$ 1,949,306,188}} & \multicolumn{3}{c|}{\textbf{1,949,306,188 $\triangleright$ 1,950,379,470}} & \multicolumn{3}{c|}{\textbf{1,950,379,470 $\triangleright$ 1,951,309,711}} & \multicolumn{3}{c}{\textbf{1,951,309,711 $\triangleright$ 1,953,752,969}} \\ \midrule
 & time & \multicolumn{1}{c}{peak} & \multicolumn{1}{c|}{static} & time & \multicolumn{1}{c}{peak} & \multicolumn{1}{c|}{static} &  time & \multicolumn{1}{c}{peak} & \multicolumn{1}{c|}{static} &  time & \multicolumn{1}{c}{peak} & \multicolumn{1}{c}{static} \\ \midrule
Standrad & \multicolumn{1}{r}{$>$38h} & \multicolumn{1}{r}{-} & \multicolumn{1}{r|}{-} & \multicolumn{1}{r}{-} & \multicolumn{1}{r}{-} & \multicolumn{1}{r|}{-} & \multicolumn{1}{r}{-} & \multicolumn{1}{r}{-} & \multicolumn{1}{r|}{-} & \multicolumn{1}{r}{-} & \multicolumn{1}{r}{-} & \multicolumn{1}{r}{-} \\
TCModule & \multicolumn{1}{r}{3295.25} & \multicolumn{1}{r}{97,940.15} & \multicolumn{1}{r|}{82,348.18} & \multicolumn{1}{r}{138.40} & \multicolumn{1}{r}{97,940.15} & \multicolumn{1}{r|}{82,101.52} & \multicolumn{1}{r}{30.04} & \multicolumn{1}{r}{97,940.15} & \multicolumn{1}{r|}{82,132.15} & \multicolumn{1}{r}{1519.99} & \multicolumn{1}{r}{98,406.95} & \multicolumn{1}{r}{82,548.45} \\
TCScheme & \multicolumn{1}{r}{443.80} & 2,447.91 & 2,304.00 & \multicolumn{1}{r}{2.22} & 2,447.91 & 2,304.84 & \multicolumn{1}{r}{2.96} & 2,447.91 & 2,304.85 & \multicolumn{1}{r}{14.84} & 2,447.91 & 2,304.87 \\ 
\bottomrule
\end{tabular}%
}
\caption{Performance Evaluation of TC Scheme Algorithms on DBpedia's \textit{broader} relation. The bold text indicates changes in the fact count before and after materialisation. 
The \textit{time} is in second, \textit{peak} and \textit{static} stand for the peak memory usage during the reasoning and the static memory used by the data structure, respectively. Both of the memory are reported in MB. }
\label{tab:testTCinc}
\end{table*}

\begin{table*}[t]
\centering
\resizebox{\textwidth}{!}{%
\begin{tabular}{l|rrrrr|rrrrr|rrrrr}
\toprule
 & \multicolumn{5}{c|}{\textbf{\begin{tabular}[c]{@{}c@{}}DB25\% (23.0M $\triangleright$ 32.7M);   \\      Union: 4.2M; TC: 677.8k $\triangleright$ 4.2M\end{tabular}}} & \multicolumn{5}{c|}{\textbf{\begin{tabular}[c]{@{}c@{}}DB50\% (46.0M $\triangleright$ 608.7M); \\      Union: 280.2M; TC: 1.4M $\triangleright$ 280.2M\end{tabular}}} & \multicolumn{5}{c}{\textbf{DAG (100k $\triangleright$ 22.9M)}} \\
 \midrule
 & time & \multicolumn{1}{c}{peak} & \multicolumn{1}{c}{static} & \multicolumn{1}{c}{$NT$} & other $T$ & \multicolumn{1}{c}{time} & \multicolumn{1}{c}{peak} & \multicolumn{1}{c}{static} & \multicolumn{1}{c}{$NT$} & \multicolumn{1}{c|}{other $T$} & \multicolumn{1}{c}{time} & \multicolumn{1}{c}{peak} & \multicolumn{1}{c}{static} & \multicolumn{1}{c}{$NT$} & other $T$ \\
 \midrule
Standard & \multicolumn{1}{l}{33.42} & 2.7 & 2.2 & 2.2 & \multicolumn{1}{c|}{-} & 12,879.90 & 30.8 & 25.4 & 25.4 & \multicolumn{1}{c|}{-} & 3,973.21 & 1.1 & 0.9 & \multicolumn{1}{c}{0.9} & \multicolumn{1}{c}{-} \\
TCModule & \multicolumn{1}{l}{33.88} & 3.2 & 2.7 & 2.2 & \multicolumn{1}{c|}{-} & 984.86 & 31.8 & 26.4 & 25.4 & \multicolumn{1}{c|}{-} & 36.04 & 1.2 & 1.0 & \multicolumn{1}{c}{0.9} & \multicolumn{1}{c}{-} \\
MultiScheme & \multicolumn{1}{l}{31.39} & 3.6 & 3.1 & 1.9 & \multicolumn{1}{c|}{1.2} & 479.35 & 8.7 & 7.7 & 3.1 & \multicolumn{1}{c|}{4.6} & 26.45 & 0.9 & 0.9 & 0.01 & \multicolumn{1}{c}{0.89} \\
\midrule
 & \multicolumn{5}{c|}{\textbf{\begin{tabular}[c]{@{}c@{}}DB75\% (69.0M $\triangleright$ 8.6B);\\      Union: 4.3B; TC: 2.0M $\triangleright$ 4.3B\end{tabular}}} & \multicolumn{5}{c|}{\textbf{\begin{tabular}[c]{@{}c@{}}DBAll (92.0M $\triangleright$ 34.9B); \\      Union: 17.4B; TC: 2.7M $\triangleright$ 17.4B\end{tabular}}} & \multicolumn{5}{c}{\textbf{\begin{tabular}[c]{@{}c@{}}Relations (845.8k $\triangleright$ 212.2M);   \\      Union: 191.0M; TC: 390.3k $\triangleright$ 14.1M\end{tabular}}} \\
 \midrule
 & time & \multicolumn{1}{c}{peak} & \multicolumn{1}{c}{static} & \multicolumn{1}{c}{$NT$} & other $T$ & \multicolumn{1}{c}{time} & \multicolumn{1}{c}{peak} & \multicolumn{1}{c}{static} & \multicolumn{1}{c}{$NT$} & \multicolumn{1}{c|}{other $T$} & \multicolumn{1}{c}{time} & \multicolumn{1}{c}{peak} & \multicolumn{1}{c}{static} & \multicolumn{1}{c}{$NT$} & other $T$ \\
 \midrule
Standard & \multicolumn{1}{r}{\textgreater{}86h} & \multicolumn{1}{c}{-} & \multicolumn{1}{c}{-} & \multicolumn{1}{c}{$\approx$515} & \multicolumn{1}{c|}{-} & \multicolumn{1}{c}{-} & \multicolumn{1}{c}{-} & \multicolumn{1}{c}{-} & \multicolumn{1}{c}{$\approx$2094.9} & \multicolumn{1}{c|}{-} & \multicolumn{1}{r}{14,277.00} & \multicolumn{1}{r}{11.0} & \multicolumn{1}{r}{9.3} & \multicolumn{1}{r}{9.2} & \multicolumn{1}{c}{-} \\
TCModule & \multicolumn{1}{r}{\textgreater{}86h} & \multicolumn{1}{c}{-} & \multicolumn{1}{c}{-} & \multicolumn{1}{c}{-} & \multicolumn{1}{c|}{-} & \multicolumn{1}{c}{-} & \multicolumn{1}{c}{-} & \multicolumn{1}{c}{-} & \multicolumn{1}{c}{-} & \multicolumn{1}{c|}{-} & \multicolumn{1}{r}{2,046.36} & \multicolumn{1}{r}{11.1} & \multicolumn{1}{r}{9.4} & \multicolumn{1}{r}{9.2} & \multicolumn{1}{c}{-} \\
MultiScheme & \multicolumn{1}{r}{5,973.34} & 16.3 & 14.9 & 4.0 & \multicolumn{1}{c|}{10.9} & 23,092.37 & 17.3 & 15.1 & \multicolumn{1}{c}{5.6} & \multicolumn{1}{c|}{9.5} & 6,905.72 & 4.9 & 4.1 & 3.6 & \multicolumn{1}{c}{0.5} \\
\bottomrule
\end{tabular}%
}
\caption{Performance Evaluation of Multi-Scheme Reasoning Algorithm. 
\textit{Time} is in seconds. The other four metrics are in GB, in which $NT$ and `other $T$' mean the memory used by the normal plain table and other schemes, respectively. }
\label{tab:TCR}
\end{table*}

\noindent \textbf{Benchmarks:}
We tested our algorithms on \textit{DAG-R}~\cite{hu2022modular}, \textit{DBpedia}~\cite{DBLP:journals/semweb/Lehmann2015DBpedia}, and \textit{Relations}~\cite{smith2007obo}. 
DAG-R is a synthetic benchmark, containing a randomly generated directed acyclic graph with 100k edges and 10k nodes and a Datalog program in which the \textit{connected} property is transitive. 
DBpedia consists of structured information from Wikipedia. The SKOS vocabulary\footnote{https://www.w3.org/TR/skos-reference/} is used to represent various Wikipedia categories. We used a Datalog subset of the SKOS RDF schema as rules for DBpedia, in which several transitive and union rules are present. 
The Relations benchmark is obtained from the Relations Ontology~\cite{smith2007obo} containing numerous biomedical ontologies. 
The converted Datalog program consists of 1307 rules in total, $33$ TC schemes and $130$ Union schemes are created according to the program. 
The original ontology does not have data associated, so we use the synthetic dataset created by \citet{hu2022modular}.

\noindent \textbf{Compared Approaches:}
We considered three approaches for the evaluation of materialisation time and memory consumption. The \textit{standard} approach applies the semina\"ive algorithm for materialisation and uses just normal tables for storage. 
The \textit{Multi-Scheme} is our proposed approach, using a plain table, \textit{TC} and \textit{Union} schemes.
The \textit{TC Module} approach proposed by~\citet{hu2022modular} applies an optimised application of TC rules, and a standard semina\"ive algorithm for the remaining rules, but only a plain table is used for storage. 
The original modular approach also proposes optimisations for other types of rules, such as chain rules. For a fair comparison, we evaluate a version where only the optimisation for transitive closure rules is enabled. 

\noindent \textbf{Test Setups: }
All of our experiments are conducted on a Dell PowerEdge R730 server with 512GB RAM and 2 Intel Xeon E5-2640 2.60GHz processors, running Fedora 33, kernel version 5.10.8.

\noindent \textbf{Performance of TC Scheme Algorithms:}
To comprehensively evaluate the performance of the proposed TC functions, we extracted two sets of \textit{broader} facts from DBpedia and created a program with a transitive rule for \textit{broader}. 
For each dataset, we inserted the facts in four rounds: the first insertion added all remaining facts (shown in the first column of Table~\ref{tab:testTCinc}), while the next three insertions each added 1,000 new facts as $E^+$ to test incremental maintenance (the last three columns).
For the smaller dataset (the upper rows), the TC Module approach optimised the running time to a large extent compared to the standard approach, but not on memory consumption.
In contrast, our TC scheme approach 
is around 100-1000x faster than the standard approach, but only uses about 1/8 $\sim$ 1/5 memory, in all the tasks. 
For the larger dataset (the lower rows), the standard approach failed to finish the initial insertion. 
Our TC scheme approach finished all the tasks and only used around 
1/35 of the memory used by the TC Module.
Our TC scheme can also maintain the data structure quickly under addition (around 7-100x faster than the TC Module), which is beneficial for the recursive and incremental reasoning scenario.  

\noindent \textbf{Performance of Multi-Scheme Reasoning Algorithms: }
We tested the performance of our proposed multi-scheme reasoning algorithm on the benchmarks mentioned above.
A scalability evaluation was conducted by randomly choosing subsets from DBpedia.
As shown in Table~\ref{tab:TCR}, our multi-scheme approach used slightly more time and memory than the standard approach when the dataset is small (\textit{DB25\%}), since the fraction between the output TC facts and the input TC facts is small, benefits of using the compressed data structure does not show. 
However, for \textit{50\%} subset of DBpedia, our approach is 27x faster than the standard approach and only uses 1/3 memory.
In the reasoning process, TC and union schemes naturally consume more time to traverse the contained facts than the plain table, which will lead to longer rule application time for the multi-scheme approach. 
But this approach is still faster than applying TC and union rules faithfully as done in the standard approach. 
The standard and TC Module approach cannot finish the materialisation for \textit{75\%} and the whole DBpedia; while our approach completes the materialisation only using around $15$ GB. 
In contrast, storing the materialisation of \textit{75\%} and full DBpedia is estimated to take $515$ GB and $2094.9$ GB respectively. 

For DAG-R, using TC schemes speeds up runtime but does not reduce memory usage by much, due to the size of the TC closure. 
For Relations, the TC rule optimisation in the TC Module significantly decreases materialisation time, but not memory usage. In contrast, our approach finishes materialisation using less than half the memory of the standard and TC Module approaches. 
However, the presence of union predicates in some rule bodies requires traversing facts represented by union schemes, increasing running time, though it is still faster than the standard approach.

\begin{table}[t]
\centering
\resizebox{\columnwidth}{!}{%
\begin{tabular}{l|rrrrr}
\toprule
TC Query & $Q_0$ (140M) & $Q_1$ (140M) & $Q_3$ (428) & $Q_7$ (752k) & $Q_8$ (1M) \\
 \midrule
Standard & 8.38 & 27.70 & 0.03 & 0.10 & 0.13 \\
MultiScheme & 22.43 & 20.62 & 0.03 & 0.24 & 0.35 \\
\midrule
Union Query & $Q_0$ (280M) & $Q_2$ (753k) & $Q_3$ (1M) & $Q_4$ (337) & $Q_8$ (1M) \\
 \midrule
Standard & 25.50 & 0.28 & 0.42 & 0.03 & 0.15 \\
MultiScheme & 338.90 & 0.61 & 0.89 & 0.03 & 2.93 \\
\bottomrule
\end{tabular}%
}
\caption{The Query Answering Time in seconds. The cardinality of each query is provided in the brackets. }
\label{tab:qa}
\end{table}

\noindent \textbf{Performance of Query Answering:}
One potential disadvantage of the multi-scheme framework is increased query retrieval time. To fully characterise this trade-off, we evaluated query performance using $11$ queries with transitive predicates and $11$ queries with union predicates. 
Instead of using queries with complex graph patterns, we employ queries with 1 or 2 atoms using the TC or union predicate to capture the performance of specialised storage schemes. 
Query execution times were conducted in \textit{50\%} subset of DBpedia and compared against the standard approach. 
Due to page limits, Table~\ref{tab:qa} presents results for 5 queries with transitive predicates in the upper rows and 5 queries with union predicates at the bottom; the complete table and all queries are provided in the \inappendix.
The evaluation results suggest that for queries with small cardinality (usually less than 1 million), the running time is not significantly different. 
For queries with the transitive predicate, our approach consumes less than 3 times of the time used by the standard approach. 
For queries with union predicates, it is around 2-20 times, since retrieval from union schemes includes querying other schemes to remove the duplicate and verify the status of related facts.

\section{Discussion and Perspectives}

In this paper, we proposed a framework that can accommodate different storage and reasoning optimisations.
Our approach offers a flexible and extensible alternative that supports Datalog reasoning applications in scenarios where storage resource is limited and materialisation fails. Future work will involve supporting deletion in the multi-scheme framework and introducing deletion functions for specialised tables. For example, maintaining the data structure in the TC scheme will require handling the collapse of an SCC caused by deletion.

\section*{Acknowledgements}
This work was supported by the following EPSRC projects: OASIS (EP/S032347/1), UK FIRES (EP/S019111/1), and ConCur (EP/V050869/1), as well as by SIRIUS Center for Scalable Data Access, Samsung Research UK, and NSFC grant No. 62206169.

\bibliography{aaai24}



\section*{Supplementary material}

# Optimised Storage for Datalog Reasoning - Appendix

## Plain Table Storage Scheme

We present the functions of the plain table in algorithm 1, in which the facts that are not managed by any specialised storage schemes are stored. The predicate set $P_T$ is defined as the predicate that is not supported by any other schemes. The $\Pi_T$ and $SP_T$ are initialised according to $P_T$. The internal data structure $D_T$ is implemented as a fact list $L_T$, in which each fact is assigned a label, either '$\Delta$', '$I$', or '$\Delta_n$'. Then $\Delta^T$, $I^T$ and $\Delta_n^T$ are defined intuitively to return all the facts with the specified label:

$$I^T = \{t \in L_T \mid \text{the label of } t \text{ is '}I\text{'}\}, \quad (1)$$

$$\Delta^T = \{t \in L_T \mid \text{the label of } t \text{ is '}\Delta\text{'}\}, \quad (2)$$

$$\Delta_n^T = \{t \in L_T \mid \text{the label of } t \text{ is '}\Delta_n\text{'}\}. \quad (3)$$

The facts in $\Delta_n^T$ can be added to the fact list directly because the addition of $\Delta_n^T$ will not imply the existence of any facts. For some specialised tables, if the addition of facts does imply other facts when being inserted into the data structure, $\Delta_n^T$ should be held until *derive* is called so that $\Delta_n^T$ and its derivations w.r.t. certain rules are ready in $\Delta^T$ at the same time. Therefore, for a fact $t$ with the predicate in $P_T$ and not present in the list, the *schedule* function adds this fact to the list with the label '$\Delta_n$'. The *schedule* also populates $C_T$ by facts with the predicate in $SP_T$ that will be used in the *derive* function to derive new facts. The *derive* function marks facts in $L_T$ with the label '$\Delta_n$' to '$\Delta$', which will make facts in $\Delta_n^T$ available in $\Delta^T$. Also, $\Pi_T$ is applied by using facts in $C_T$ as new facts, derivations are added to $L_T$ as '$\Delta$' if they are not present in $L_T$. Please note that $\Pi_T$ is initialised as rules that use predicate in $P_T$ as heads, so derivations of $\Pi_T$ can be managed by $T$ itself. The *merge* function is realised by simply changing the label of facts.

It is easy to verify that the plain table is a valid scheme as defined in the main submission since *derive* and *merge* functions have the following property:

**Lemma 1** *The derive function of the plain table updates $\Delta^T$ as below, which satisfies the definition of a scheme derive function by reaching the lower bound of $\Pi_T^+[I, C_T]$:*

$$\Delta^T = \Delta_n^T \cup \Pi_T[I, C_T] \setminus I. \quad (4)$$

Algorithm 1: Functions of Plain Table
---
1: $T$: a plain table; $t$: a fact; $L_T$: the fact list in $T$.
2: $D_T$: implemented as a fact list $L_T$.
3: **procedure** $T.schedule(t)$
4:    **if** $t.p \in P_T, t \notin L_T$ **then**
5:       $L_T := L_T \cup \{t\}$
6:       the label of $t :=$ '$\Delta_n$'
7:    **if** $t.p \in SP_T$ and $t$ has not been seen by $T$ **then**
8:       $C_T := C_T \cup \{t\}$
9: **procedure** $T.merge()$
10:    **for** $t \in L_T$ **do**
11:       **if** label of $t =$ '$\Delta$' **then** label of $t :=$ '$I$'
12: **procedure** $T.derive()$
13:    mark facts in $L_T$ with '$\Delta_n$' as '$\Delta$'
14:    add facts in $\Pi_T[I, C_T]\setminus L_T$ to $L_T$ as '$\Delta$'
15:    $C_T = \emptyset$

**Lemma 2** *The merge function of the plain table updates $I^T$ as $I^T \cup \Delta^T$ using values before merging, and empties $\Delta^T$.*

## TC Storage Scheme

We first present the complete version of functions in the TC storage scheme, and then prove that functions of this scheme follow the property stated in the main submission. We define the transitive closure scheme that can compute the closure w.r.t. the input facts internally and store facts compactly in algorithm 2. Assuming there is a rule $r$ that axiomatises a relation $R$ as transitive; the predicate set $P_T$ contains $R$. Then rule $r$ is included in $\Pi_T$, as shown in line 2. For the sake of this presentation, we assume $\Pi_T$ only contains $r$. Therefore, the predicate set $SP_T$ only contains the predicate $R$. The facts with the predicate $R$ that are scheduled to the current scheme $T$ are stored in both $\Delta_n^T$ and $C_T$, since these facts are new to the scheme ($\Delta_n^T$) and also can be used to derive new facts with the same predicate $R$. For brevity, we only use the set $C_T$ in algorithm 2. During the *schedule* function, the fact with the predicate $p \in P_T$ is added to the cache of the TC scheme, if the fact is not represented in the data structure $D_T$ nor stored in $C_T$. The *derive* function inserts $C_T$ to the data structure such that $C_T$, and the derivations derived because of the inserting of $C_T$, are available in $\Delta^T$.

```
Algorithm 2: Functions of TC storage scheme
1:  T: a TC storage scheme; t : a fact; D_T: G, L as in text.
2:  P_T : {R}. Π_T = {R(x,z), R(z,y) → R(x,y) ∈ Π}.
3:  procedure T.schedule(t)
4:      if t.p ∈ P_T, t ∉ C_T and t ∉ I^T ∪ Δ^T then
5:          C_T := C_T ∪ {t}
6:  procedure T.merge()
7:      for s ∈ L do                         ▷ G is updated in Add
8:          if s.St = dropped then delete s from L
9:          else s.In += s.DIn, s.DIn = s.NIn = ∅
10:             if s.St = new then s.St = stable
11:                 M(s) = nM(s), F(s) = nM(s) = ∅
12: procedure T.derive()
13:     if D_T has not been initialised then
14:         TC_Initialisation(C_T)           ▷ algorithm 3
15:     else TC_Add(C_T)                     ▷ algorithm 4
16:     C_T = ∅
```

In this section, we only discuss the transitive closure rule, all the other rules included in $\Pi_T$ are processed as the plain table, but the derivations are managed by the TC scheme.

## The Data Structure of TC Storage Scheme

The data structure $D_T$ of a TC storage scheme is realised by an adjacency list $G$, and an ordered list $L$. The $G$ stores the graph representation of input facts, based on which the closure is computed. The list $L$ is a collection of nodes in $G$, ordered by the post-order index. Recall the example in the main text, for each node in $L$, the post-order index, associated intervals, and other relevant information are stored.

## TC Scheme Initialisation

When *derive* is called the first time and $D_T$ has not been initialised, algorithm 3 is called to initially construct the data structure based on the facts scheduled for insertion in $C_T$. The example presented in figure 1 shows the process of TC storage scheme initialisation. The initial construction is basically the same as the algorithm introduced by Agrawal, Borgida, and Jagadish (1989), but we introduce a new data structure $L$ to support new functions we introduce for the TC scheme, i.e., the maintenance of $D_T$ under addition and the algorithm to access $I^T$ and $\Delta^T$. The input fact set $TS$ is first represented as a graph $G_0$ (line 3). Then, $G_0$ is condensed to $G$ by shrinking each strongly connected component (i.e., SCC) as a single node in line 4. The nodes $v \in V$ of $G$ are a disjoint set of nodes in $V'$ of $G_0$, we call $v \in V$ an SCC, and $v \subseteq V'$. This step is to convert a possibly cyclic graph $G_0$ to a directed acyclic graph $G$. Disconnected subgraphs in $G_0$ are connected by introducing a "*fake root*" and adding edges between the fake root and the root of each subgraph in line 7. After this step, one tree cover $Tree$ is computed to cover all the nodes in $G$. Finally, the post-order index and intervals are computed for each SCC. The following members are initialised for each SCC $s$ in $L$:

1. The post-order index of the SCC $s$: $s.id$.
2. The map $M$ maps an SCC $s$ to its contained resources.
3. The status $St$ of $s$ is used to identify whether $s$ is "*stable*", newly merged from existing SCCs due to fact insertion ("*new*"), or merged to a *new* SCC ("*dropped*"). In the initialisation, the statuses of all SCCs are *stable* since no merging happened between these SCCs.
4. The intervals $In, DIn, NIn$ includes the indexes of SCCs that $s$ can reach in different labels. Intuitively, $In$ records the SCCs that $s$ can reach in '$I$', $DIn$ and $NIn$ are for '$\Delta$'. If $s$ is *stable*, $s.DIn$ is used to memorise "new" intervals (different from $In$) that include the indexes of SCCs that $s$ can reach after fact addition, so $s.DIn \nsubseteq s.In$. But when $s$ is *new*, $s.DIn \nsubseteq s.In$ is not true. The $s.DIn$ is the interval that includes the indexes of SCCs that $s$ can reach after merging, regardless of labels. For any type of SCC, the $NIn$ stores the interval that includes new descendants of $s$ introduced via addition, so $s.NIn \subseteq s.In \cup s.DIn$. If facts with the label '$I$' is needed, SCCs in $s.NIn \cap s.In$ need to be skipped.
5. The tree interval $TIn$ includes the indexes of SCCs located in the subtree rooted at $s$.
6. The map $nM$ maps an SCC $s$ to its updated contained constants if $s$ is newly merged and $s.St$ = *new*.
7. The map $F$ maps an SCC $s$ to a set of SCC, from which $s$ is merged. The map $nM$ and $F$ are useful when the fact addition causes existing SCCs to merge as one.

During the initialisation, the $In$ of all the SCCs is empty, $DIn$ and $TIn$ are computed according to the tree cover and graph edges in $G$. The $NIn$, $nM$, and $F$ are designed for incremental addition and not used in the initialisation.

## Facts in $\Delta^T$ and $I^T$

Define $\{A \times B\}$ as a set of facts in which $A$ is a set of subjects, while $B$ is a set of objects, the predicate is always $R \in P_T$, and the operator $\times$ means the cross-product between the resources. We define operator $H$ to return the up-to-date version of contained resources for an SCC $s$:

$$H(s) = \begin{cases} M(s), & s \text{ is } stable \\ nM(s), & s \text{ is } new \\ \emptyset, & s \text{ is } dropped \end{cases} \quad (5)$$

For the ordered collection $L$, an instantiation of an interval $In$ w.r.t. $L$ is defined as:

$$L_{In} = \{s \in L \mid s.id \in In\} \quad (6)$$

The $M$ and $H$ operators can also be generalised by taking an interval as input:

$$M(In) = \bigcup_{s \in L_{In}} \{M(s)\}, \quad H(In) = \bigcup_{s \in L_{In}} \{H(s)\}. \quad (7)$$

For a TC storage scheme $T$, the facts that can be returned under different domains are defined as follows, in which $I_s$, $\Delta_s$, $I_s \cup \Delta_s$ are defined in table 1.

$$I^T = \bigcup_{s \in L} \{I_s\}, \quad \Delta^T = \bigcup_{s \in L} \{\Delta_s\} \quad (8)$$

$$I^T \cup \Delta^T = \bigcup_{s \in L} \{I_s \cup \Delta_s\} \quad (9)$$

| status | stable | new |
|---|---|---|
| $I_s$ | $\{M(s) \times M(s.In \setminus s.NIn)\}$ | $\bigcup_{n \in F(s)} \{M(n) \times M(n.In \setminus s.NIn)\}$ |
| $\Delta_s$ | $\{M(s) \times M(s.DIn)\} \cup \{M(s) \times M(s.NIn \setminus s.DIn)\}$ | $\bigcup_{n \in F(s)} (\{M(n) \times M(s.DIn \setminus n.In)\} \cup \{M(n) \times M(s.NIn \setminus (s.DIn \setminus n.In))\})$ |
| $I_s \cup \Delta_s$ | $\{M(s) \times H(s.In + s.DIn)\}$ | $\{nM(s) \times H(s.DIn)\}$ |

Table 1: For an SCC $s$, the generated facts in different domains, represented as the cross product '×' between a set of subjects and a set of objects. The predicate is omitted. If $s.St = dropped$, $I_s = \Delta_s = I_s \cup \Delta_s = \emptyset$.

For a *stable* $s$, for all domains, $M(s)$ is unchanged and used as subjects. While the objects for $I_s$ are found in the nodes of $L_{s.In}$, but skipping nodes in $L_{s.NIn}$. Because the nodes in $L_{s.NIn}$ are newly introduced and can never be found in $I_s$. Case 3 introduced below includes one such example. For $\Delta_s$, the objects will include the "new" intervals assigned during addition, i.e., $s.DIn$. Additionally, the nodes that are newly inserted (i.e., $s.NIn \cap s.In$ or $s.NIn \setminus s.DIn$ as shown in the table) should also be included. For $I_s \cup \Delta_s$, the objects will be the union of the objects of $I_s$ and $\Delta_s$, i.e., $M(s.In + s.DIn) = H(s.In + s.DIn)$[1].

If the fact addition causes several SCCs $C \subseteq V$ to merge, one of these SCCs (let it be $s \in C$) will be chosen as representative and the status is *new*, and the left SCCs are *dropped*. To retrieve $I^T$ and $\Delta^T$, the dropped SCCs are not deleted before the call of *merge*. For each $n \in C$, the $n.In$ and $M(n)$ stay the same as before addition, representing the interval and contained constants of $n$ before merging. For $s$, the $s.DIn$ is the interval containing all the nodes that $s$ can reach after merging, the formal expression is shown in line 14 of algorithm 4. Similar for $s.NIn$ (line 15). Please note that for *new* $s$, $s.DIn \not\subseteq s.In$ is not true (different from *stable* SCC), but $s.NIn \subseteq s.In \cup s.DIn$ is still true. And $F(s)$ contains the SCCs from which the new $s$ is merged. The $nM(s)$ contains the union of constants of $n \in C$. Please refer to Case 2 below for more discussions about how to maintain the data structure when SCC is changed by addition.

For a *new* $s$, for $I_s$ and $\Delta_s$, the "old" SCCs from which $s$ is merged, i.e., $F(s)$ is traversed. The old constants in $n \in F(s)$, i.e., $M(n)$, are used for subjects. For the object of $I_s$, the old interval of $n$ but skipping the new nodes in $s.NIn$ is used, i.e., $n.In \setminus s.NIn$. While for the object of $\Delta_s$, $n$ can reach all the nodes that $s$ can reach after merging, i.e. $s.DIn$, but only the nodes in $s.DIn \setminus n.In$ is new. Additionally, all the newly introduced nodes in $s.NIn$ should be included in $\Delta_s$, but leaving the interval $s.DIn \setminus n.In$ to avoid duplication. For $I_s \cup \Delta_s$, we can use the $s$ after merging directly. The subject is the updated constants in $s$, i.e., $nM(s)$. The object is updated interval for $s$ after merging, i.e., $s.DIn$. The expression shown in table 1 is equivalent to the pure union of $I_s$ and $\Delta_s$.

### Incremental Maintenance Algorithm

We present the complete *TC_add* function in algorithm 4. To simplify the pseudo code, we defined a function *initialiseSCC* in algorithm 5 to initialise the members of an SCC if the current edge to be inserted contains fresh constants.

---
[1] The equivalence will be proved in appendix.

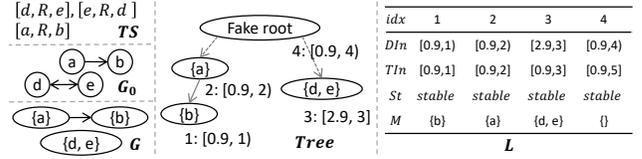

Figure 1: One Example of TC storage scheme Initialisation.

---

Algorithm 3: TC_Initialisation

1: **Input:** $TS$, a fact set.
2: **Result:** construct $G, L$ in $T$
3: $G_0(V', E') = buildGraph(TS)$
4: $G(V, E) = graphCondensation(G)$  ▷ $G$ is a DAG
5: add a fake root $f$ to $V$
6: add $\{(f, v) \mid v \in V, inDegree(G, v) = 0, v \neq f\}$ to $E$
7: $Tree = treeCover(G)$  ▷ $Tree \subseteq E$
8: add $v \in V$ to $L$ in post order of $Tree$
9: assign intervals to $v \in V$ and store in $L$

---

This function initialises a new strongly connected component (SCC) $s$ as a child node of the existing node $p$ in the graph $G$ maintained in the TC storage scheme. In line 3, the *findGap* function will find an empty branch in the subtree rooted at $p$ and assign this empty branch as the tree interval of $s$. Formally, the empty branch assigned to $s$ is a subset of the intervals that belong to the tree interval of the parent $p$, but are not assigned to any child of $p$ yet, i.e., $s.TIn \subseteq (p.TIn \setminus \bigcup_{c \in \text{child}(p)} \{c.TIn\}$. Also, $s.DIn$ is initialised according to the $s.TIn$, since every SCC belongs to $s.TIn$ is a new SCC and should present in '$\Delta$'. The difference between $s.DIn$ is $s.TIn$ is that $s.TIn$ that represents the indexes of nodes rooted at $s$ includes $s$ itself, but $s.DIn$ does not include $s$ unless the edge $(s, s)$ is explicitly inserted or the SCC $s$ contains more than one constant. The $s.In$ is initialised as empty in line 3, because $s$ is newly introduced, nothing is in '$I$'. In line 4, the post-order index is initialised as the upper bound of the tree interval. Finally, the status of $s$ is assigned as *stable* since $s$ is not merged from several SCCs.

The incremental maintenance of the data structure in the TC storage scheme will be processed edge by edge in line 3 of algorithm 4. For each edge $(i', j')$ in which $i'$ and $j'$ are constants (e.g., using the example in figure 1, assume one edge between $b$ and $e$ is introduced), the corresponding SCC in $V$ needs to be found first in line 4. In the above example, the SCC $i$ in which $i'$ is contained is with the index 1; while for $j'$, it is with the index 3. If an existing $SCC$ containing such a constant (let it be $i'$) cannot be found, a new

## Algorithm 4: TC_add

1: **Input:** $TS$, a fact set
2: **Result:** update $G$ and $L$ s.t. $I^T, \Delta^T$ is correct
3: **for** $(i', j') \in TS$ **do**  ▷ predicate is omitted here
4:     $i, j = findSCC(i', j')$
5:     /* **Case 1:** [non-tree] cross edge */
6:     **if** $i, j \in L$ and $(i', j'), (j', i') \notin I^T \cup \Delta^T$ **then**
7:         $i.DIn\ += j.In + [j.idx] + j.DIn$
8:         **if** $i.St = stable$ **then** $i.DIn\ -= i.In$
9:     /* **Case 2:** [non-tree] back edge */
10:    **if** $i, j \in L, (j', i') \in$, but $(i', j') \notin I^T \cup \Delta^T$ **then**
11:        $C = findSCCToMerge(i, j)$  ▷ $C \subseteq V$ of $G$
12:        $nM(i)\ += \bigcup_{n \in C}\{H(n)\}$
13:        $F(i)\ += \bigcup_{n \in C}\{F(n)$ if $F(n) \neq \emptyset$ else $\{n\}\}$
14:        $i.DIn\ += \bigcup_{n \in C}\{n.In + n.DIn + [n.idx]\}$
15:        $i.NIn\ += \bigcup_{n \in C}\{n.NIn\}$
16:        $i.TIn\ += \bigcup_{n \in C}\{n.TIn\}$
17:        $i.St = new$, $n.St = dropped$ for $n \in C$
18:    /* **Case 3:** [tree-edge] new child */
19:    **if** $i \in L, j \notin L$ **then**
20:        $initialiseSCC(j, i), i.NIn\ += j.TIn$
21:        add $j$ to $L$  ▷ position decided by $j.idx$
22:    /* **Case 4:** [tree-edge] new parent */
23:    **if** $i \notin L, j \in L$ **then**
24:        $initialiseSCC(i, f)$  ▷ $f$ is the fake root
25:        $i.DIn\ += j.In + j.DIn + [j.idx]$
26:        add $i$ to $L$
27:    /* **Case 5:** [tree-edge] new child and parent */
28:    **if** $i \notin L, j \notin L$ **then**
29:        $initialiseSCC(i, f), i.NIn\ += i.DIn$, add $i$ to $L$
30:        **if** $j \neq i$ **then**
31:            $initialiseSCC(j, i)$, add $j$ to $L$
32:    /* **Case 6:** [non-tree] forward edge */
33:    **if** $(i', j') \in I^T \cup \Delta^T$ **then** continue
34:    update $G$  ▷ include merging the SCCs
35:    $propagate(i)$  ▷ algorithm 6

## Algorithm 5: initialiseSCC

1: **Input:** $s$, an SCC to be initialised; $p$, the parent of $s$.
2: **Result:** initialise the intervals, and other members of $s$.
3: $s.DIn, s.TIn = findGap(p), s.In = \emptyset$
4: $s.idx =$ the upper bound of $s.TIn$
5: $s.St = stable$

## Algorithm 6: propagate

1: **Input:** an SCC $n \in G$
2: **Result:** pass $n.NIn$ and $n.DIn$ to the ancestors of $n$
3: **for** $m \in father(G, n)$ **do**
4:     **if** $n.DIn \subseteq m.DIn + m.In$ and $n.NIn \subseteq m.NIn$ **then** continue
5:     **if** $m.St = stable$ **then**
6:         $m.DIn\ += n.DIn \setminus m.In$
7:     **else** $m.DIn\ += n.DIn$  ▷ $m.St = new$
8:     $m.NIn\ += n.NIn$
9:     $propagate(m)$

SCC $i$ consisting of a single item $i'$ will be created. Each edge inserted is categorised into one of several cases, and then processed differently depending on which case it falls into. The cases 1-2 and 6 are about the "non-tree" edges, in which $i$ and $j$ exist in $L$ already and the insertion will not change the tree cover of $G$, but only introduce new connections between existing SCCs. Then according to whether $i'$ and $j'$ can reach each other, this edge will be classified to a specific case: if $i'$ can reach $j'$ already, then there is nothing needs to be done (case 6); If neither $i'$ nor $j'$ can reach each other, this edge is a cross edge, and after insertion, $i'$ can reach everything that $j'$ can reach (case 1); If $i'$ cannot reach $j'$, but $j'$ can reach $i'$, then the insertion of $(i', j')$ will cause the merging of several SCCs (case 2). The reachability between $i'$ and $j'$ is identified by $I^T \cup \Delta^T$ defined in the expression 9. Then we introduce case 1 and case 2 in detail below.

**Case 1:** The insertion of $(b, e)$ will not change the tree cover, but introduce a *cross edge* between two branches. Formally, the condition to identify this case is described in line 5, neither $b$ nor $e$ can reach each other. In this case, if $i$ can reach $j$ after addition, then $i$ can reach the nodes that $j$ can reach. Therefore, the index of $j$, as well as $j.In + j.DIn$, is added to $i.DIn$. The $DIn$ represents the "delta" intervals that the current node cannot reach before the insertion of this edge, but can reach after the current insertion serves as a "bridge" between $i$ and $j$.

**Case 2:** Another case of non-tree edge is a *back edge* (line 9), in which several SCCs in $G$ will be merged. In this case, a set $C$ of current nodes in $G$ is identified first in line 11. Assume we choose $i$ as the representative, and the $i$ after merging will (1) contain all the constants of nodes in $C$ (line 12); (2) all the nodes that $C$ can reach, $i$ can reach after merging (line 14 - 15); (3) the subtree in which $i$ is the root will be the "union" of the subtrees of nodes in $C$ (line 16). To keep the original $I^T$, instead of updating the $M$ and $In$ directly, we use another map $nM$ to memorise the updated constants and use $DIn$ to store temporary new intervals. In line 12, $nM(i)$ will have the most updated constants for every $n \in C$ by using the operator $H$. For every $n \in C$, the *merged from* set $F(i)$ includes $F(n)$ if $n$ has been merged before, else includes $n$. Then the status of the nodes in $C$ will be updated in line 17.

The remaining cases 3-5 are about the "tree" edge. If $j$ is a new node introduced for an existing node $i$ (case 3), $j$ is first initialised as a child of $i$ in line 20 and $j.TIn$ is added to $i.NIn$ since the nodes in the subtree rooted at $j$ will be newly inserted. If $i$ is a new parent for $j$ (case 4), then $i$ is initialised as a child of the fake root in line 24. The edge $(f, i)$ is included in the tree cover while $(i, j)$ is processed as a cross edge. In line 25, everything $j$ can reach, as well as $j$ itself, can be accessed by $i$. Therefore, $j.In + j.DIn + [j.idx]$ is added to $i.DIn$. If $i$ and $j$ are both newly introduced, $i$ is first initialised as a child of $f$, and $i.DIn$ is added to $i.NIn$ since $L_{i.DIn}$, the SCCs whose index is included in the interval $i.DIn$, is empty before the current insertion, everything

in $i.DIn$ is newly introduced. Then if $j$ and $i$ are different SCCs containing different constants (line 30), $j$ needs to be inserted separately as a child of $i$.

**Case 3**: We discuss a case in which the tree cover is changed by the fact insertion. For example, consider inserting $(e, x)$ in the example of figure 1. In line 4, $i$ is the SCC with the index 3, and we cannot find an SCC $j \in L$ that includes $x$. Let $j$ be a new SCC containing the constant $x$. Since $x$ is introducing a new subtree that is included in the subtree of $i$ and doesn't cross the other branches of $i$, the $j.TIn$ will be found in the gaps between $i.TIn \setminus \bigcup_{n \in child(i)} \{n.TIn\}$. Assume the assigned tree interval $j.TIn = [2.95, 2.98]$, then $j.DIn = [2.95, 2.98)$. The index of $j$ will be the upper bound of $j.TIn$, i.e., 2.98. In the end, $i.NIn$ is updated by adding the interval assigned to $j$ in line 20. Since $L_{j.TIn} = \emptyset$ before addition, any nodes in $L_{j.TIn}$ is newly introduced during this addition and should be included in $NIn$.

At the end of each case: (1) $G$ is updated correspondingly in line 34. Different from $L$, $G$ is updated up to date directly, since $G$ needs to reflect the latest ancestor-descendant relationships that will be used in the *propagate* function; (2) The *propagate*($i$) function is used to pass the $i.DIn$ and $i.NIn$ to $i$'s ancestors in $G$, since they can also reach what $i$ can reach after addition. In algorithm 6, for every father of $n$ in $G$, if the intervals of $n$ are already included in $m$ (line 4), then there is no need to continue. If $m$ is *stable*, then $n.DIn \setminus m.In$ will be added to $m.In$. On the other hand, if $m$ is *new* ($G$ is up to date so $m$ cannot be *dropped*), only passing $n.DIn$ is enough (line 7). Then in line 8, the $n.NIn$ is faithfully passing to $m.NIn$, since the newly introduced nodes for $n$ will also be new for $m$.

## Proof of the Properties

We first show the property that TC scheme functions satisfy, then show this property is exactly what is defined in the main submission. Normally, the order of $T$'s functions being called is fairly regular: first, the *schedule* function is called to prepare the facts to be inserted; then the *derive* function is called to insert facts and maintain the data structure; the following functions are multiple calls for $I^T$ and $\Delta^T$ that will use the current '$I$' and '$\Delta$' facts for rule application; finally *merge* is called to merge the current '$\Delta$' facts to '$I$'. The pattern above will be repeated multiple times during the whole reasoning process, which can be observed in algorithm 2 of the main submission. Recall that the *derive* function uses *TC_Initialisation* function for initialisation and uses *TC_add* function for incremental maintenance. Table 2 defines the ground truth value of $I^T$ and $\Delta^T$ after executing the first few *derive* and *merge* functions. We omit *schedule* here since it will not influence $I^T$ and $\Delta^T$. Let $r \in \Pi_T$ be the transitive closure rule managed in $T$, the *derive* will only need to compute the closure of rule $r$, since other rules in $\Pi_T$ have managed by the *schedule* function by translating the facts with the supporting predicate $SP_T$ to the transitive predicate $R$. Without loss of generality, we will show the property of the TC scheme functions by proving the function calls presented in Table 2 can maintain the data structure correctly so that $I^T$ and $\Delta^T$ contain expected facts.

**Claim 1** *The TC_initialisation(N) function call can construct the data structure correctly so that $I^T$ returns nothing, $\Delta^T$ and $I^T \cup \Delta^T$ returns $r^\infty(N)$.*

**Proof.** The *TC_initialisation* function is defined exactly the same as presented in (Agrawal, Borgida, and Jagadish 1989), the only difference is that we put the assigned intervals for each SCC $s$ in $s.DIn$. The correctness of *TC_initialisation* is based on the proof presented by Agrawal, Borgida, and Jagadish (1989). Every SCC $s \in L$ is initialised as *stable*, and $s.NIn = \emptyset$. Refer to Table 1, for any SCC $s$, $I^T$ is empty since $s.In = \emptyset$; while $\Delta^T$ can recover $r^\infty(N)$ correctly as $s.DIn$ is assigned as described in (Agrawal, Borgida, and Jagadish 1989); $I^T \cup \Delta^T$ using $s.In + s.DIn = s.DIn$ is also correct as defined in Table 2. $\square$

**Claim 2** *The merge with the index 2 in Table 2 can maintain the data structure correctly so that $I^T$ and $I^T \cup \Delta^T$ return $r^\infty(N)$, and $\Delta^T$ returns nothing.*

**Proof.** Every SCC $s \in L$ is *stable*, $s.In = s.In + s.DIn = s.DIn$, so $I^T$ recovered by using $s.In$ is the previous $\Delta^T$, i.e., $r^\infty(N)$. Additionally, $s.DIn$ and $s.NIn$ is emptied, so $\Delta^T$ is empty; while $I^T \cup \Delta^T$ using $s.In + s.DIn = s.In$ stays the same as $I^T$. $\square$

Before proving the correctness of the *TC_add* function, we first prove several lemmas about the merging of several SCCs, which will happen if a back edge is inserted.

**Lemma 3** *If the insertion of a back edge will lead to the merging of $\{n_1, ..., n_m\}$, let $s \in \{n_1, ..., n_m\}$ be a representative node, and $s.St$ = new, the others are 'dropped'. Then for every $n \in F(s)$, $n.idx \in s.In + s.DIn$. For every ancestor $a$ of $s$ in $G$, $n.idx \in a.In + a.DIn$.*

**Proof.** The $DIn$ of the representative node $s$ is used to store the intervals that $s$ can reach after merging, i.e., the union of intervals that $n \in F(s)$ can reach, as well as $n$ itself, as shown in line 14 of algorithm 4. For every $n \in F(s)$, $n.idx$ is added to $s.DIn$, so $n.idx \in s.In + s.DIn$. Then in line 35, the *propagate* function will pass $s.DIn$ to the parents of $s$ recursively until every ancestor of $s$ receives. If a parent $a$ of $s$ is *stable*, $s.DIn \setminus a.In$ is added to $a.DIn$, so $n.idx \in a.In + a.DIn$. For the part that is in $a.In \cap s.DIn$, for every ancestor $a'$ of $a$, $a.In \subseteq a'.In$, so it is safe for the call *propagate*($a$) to only pass $s.DIn \setminus a.In$ to ancestors of $a$, and it still satisfies $n.idx \in a'.In + a'.DIn$. If the parent $a.St$ = new, $s.DIn$ is added to $a.DIn$ directly. As stated above, the execution of *propagate* function will make $n.idx \in a.In + a.DIn$ for every ancestor $a$. $\square$

**Lemma 4** *For $s$ that $s.St$ = new, $n_1, n_2 \in F(s)$, we have:*

$$n_1.In + s.DIn = n_2.In + s.DIn = ID_s, \quad (10)$$
$$s.NIn \subseteq ID_s. \quad (11)$$

**Proof.** This lemma could be intuitively proved considering how $s.DIn$ and $s.NIn$ is maintained in line 14 and 15. $\square$

**Lemma 5** *For $s$ that $s.St$ = new,*

$$nM(s) = \bigcup_{n \in F(s)} \{M(n)\} = H(s) = \bigcup_{n \in F(s)} \{H(n)\}. \quad (12)$$

| idx | function call | $I^T$ | $\Delta^T$ | $I^T \cup \Delta^T$ |
|---|---|---|---|---|
| 0 | before the first call | $\emptyset$ | $\emptyset$ | $\emptyset$ |
| 1 | $TC\_initialisation(N)$ | $\emptyset$ | $r^\infty(N)$ | $r^\infty(N)$ |
| 2 | *merge* | $r^\infty(N)$ | $\emptyset$ | $r^\infty(N)$ |
| 3 | $TC\_add(\Delta)$ | $r^\infty(N)$ | $r^\infty(N \cup \Delta) \backslash r^\infty(N)$ | $r^\infty(N \cup \Delta)$ |
| 4 | *merge* | $r^\infty(N \cup \Delta)$ | $\emptyset$ | $r^\infty(N \cup \Delta)$ |
| ⋮ | ⋮ | ⋮ | ⋮ | ⋮ |

Table 2: The ground truth of $I^T$ and $\Delta^T$ of the TC storage scheme after the execution of *derive* and *merge* functions.

*Let the interval $In$ be an interval that for every $s \in L_{In}$, if $s.St = new$, then $F(s) \subseteq L_{In}$; and for every $n \in L_{In}$ if $n.St = dropped$, there exists an SCC $s \in L_{In}$ that $s.St = new$, and $n \in F(s)$. Then for $In$, we have:*

$$H(In) = M(In). \quad (13)$$

**Proof.** The $nM(s)$ and $F(s)$ are maintained in line 12 and 13, respectively. Combining with how the operator $H$ is defined in expression 5, we have $nM(s) = \bigcup_{n \in F(s)}\{M(s)\}$. Intuitively, $nM(s)$ stores the union of the constants of every SCC from which $s$ is merged from. The merged from map $F(s)$ tolerates recursion: for example, if $a$ is merged from $b$, and $b$ is merged from $c$, then $a$ can be considered as merging from $b$ and $c$ at once, i.e., $\{b, c\} \subseteq F(a)$. Then, since every $n \in F(s)$ except $s$ is *dropped*, $H(n) = \emptyset$; while for $s$, $H(s) = nM(s)$, so we have $nM(s) = H(s) = \bigcup_{n \in F(s)}\{H(n)\}$. Additionally, if we have an interval $In$ defined above, for every $s \in L_{In}$ and $s.St = new$, $\bigcup_{n \in F(s)}\{H(n)\} = \bigcup_{n \in F(s)}\{M(n)\}$. And the definition of $In$ does not allow extra dropped SCCs $n$ that there is no $s \in L_{In}$ s.t. $s.St = new$ and $F(s) = n$. For *stable* SCCs $n \in L_{In}$, $H(n) = M(n)$. As stated above, for such an interval $In$, we have $H(In) = M(In)$. □

**Lemma 6** *For an SCC $s$ that $s.St = new$, for every ancestor $a$ of $s$ in $G$, $H(a.In + a.DIn) = M(a.In + a.DIn)$.*

**Proof.** Lemma 6 can be proved by showing $a.In + a.DIn$ (let it be $In$ for simplicity) is consistent with the definition of intervals in lemma 5: (1) for every $s' \in L_{In}$, if $s'.St = new$, then $F(s') \subseteq L_{In}$, this can be shown by lemma 3. (2) for every $n \in L_{In}$ if $n.St = dropped$, there exists an SCC $s' \in L_{In}$ that $s'.St = new$, and $n \in F(s')$. This can be proved by contradiction: assume there exists an SCC $x \in L_{In}$ that $x.St = dropped$, and let $y$ be the SCC that is merged from $x$ but not in $L_{In}$, i.e., $x \in F(y)$, $y.St = new$, and $y \notin L_{In}$. Since $y.idx \notin a.In + a.DIn$, $a$ is not an ancestor of $y$ in $G$ after merging. The ancestors of $y$ after merging is the union of the ancestors of $F(y)$, so $a$ is not an ancestor of $x \in F(y)$, which is contradicted with $x \in L_{In}$. □

The above lemma can be expanded easily as follows:

**Lemma 7** *For an SCC $s \in L$ that $s.St = new$ or *stable*, $H(s.In + s.DIn + [s.idx]) = M(s.In + s.DIn + [s.idx])$.*

**Proof.** If $s$ is stable, then $H([s.idx]) = H(s) = M(s) = M([s.idx])$. If $s$ is *new*, $[s.idx] \subseteq s.In + s.DIn$. Therefore $H(s.In + s.DIn + [s.idx]) = H(s.In + s.DIn) = M(s.In + s.DIn) = M(s.In + s.DIn + [s.idx])$. □

Then, we prove the *TC_add* can maintain the data structure correctly. This function is processed edge by edge, so instead of proving the insertion of a fact set is correct, we prove the insertion of a single edge $e$ is correct, regardless of the cases. During the call of *TC_add($\Delta$)*, let $e \in \Delta$ be the single edge to be inserted next. Let $G_1, \Delta_1$ be the graph and the edges that have been processed so far (before $e$), and $G_2$ be the graph that includes $e$ and $\Delta_2 = \Delta_1 + \{e\}$. In addition, we use $I^{T_1}$ and $I^{T_2}$ to represent the facts returned before and after the insertion of $e$ in the domain '$I$', respectively. The same for the domain '$\Delta$' and '$I \cup \Delta$'. We have the following lemma:

**Lemma 8** *Assume $I^{T_1} = r^\infty(N) = I$, $\Delta^{T_1} = r^\infty[N \cup \Delta_1] \backslash I$, $I^{T_1} \cup \Delta^{T_1} = r^\infty(N \cup \Delta_1)$. The execution of the $TC\_add(\Delta_1)$ function for a single edge $e$ can update the data structure correctly so that $I^{T_2}$ stay unchanged, $\Delta^{T_2}$ returns $r^\infty(N \cup \Delta_2) \backslash I$, $I^{T_2} \cup \Delta^{T_2} = r^\infty(N \cup \Delta_2)$.*

**Proof.** We prove this lemma by proving each case presented in algorithm 4 can be maintained correctly.

- **Case 1: [non-tree] cross-edge:**

  (1) $I^{T_2} = I$: for every $s \in L$, $s.In$ and $s.NIn$ is not changed. Therefore the generated facts by Table 1 are the same as $T_1('I')$.

  (2) $\Delta^{T_2} = r^\infty(N \cup \Delta_2) \backslash I$: the insertion of $(i', j')$ builds the connection between $i'$ and $j'$, and between the ancestors of $i'$ and descendants of $j'$. Therefore, $i'$, and the ancestors of $i'$, should reach $j'$ and the descendants of $j'$ in $\Delta^{T_2}$, if it cannot reach in $I^{T_1}$. Since $I^{T_1}$ and $\Delta^{T_1}$ is correct (as a pre-condition), we use the expression of $T_1$ to have the formal definition below. In $I^{T_1} \cup \Delta^{T_1}$, the constants can be reached by $j'$, including $j'$ itself is $Y = H(j.In + j.DIn + [j.idx])$. Define the set containing the resource $i'$ and the resources that could reach $i'$ as $X = \{i'\} \cup \{x \mid (x, i') \in I^{T_1} \cup \Delta^{T_1}\}$. A set of SCCs from which the constants in $X$ can be accessed is defined as $SCC_X = \{s \mid s \in L, H(s) \subseteq X\}$. For $s \in L, s \notin SCC_X$, the *propagate* function will not change $s.DIn$ and $s.NIn$, therefore, $\Delta_s$ stays the same as $\Delta^{T_1}$, just as we expected. For $s \in SCC_X$, $s$ can be accessed by the *propagate* function to potentially maintain the intervals $s.N$ and $s.D$.

  For $s \in SCC_X$ and $s.St = stable$, $\{M(s) \times M(s.In \setminus s.NIn)\} \subseteq I^{T_1}$, and should not appear in $\Delta^{T_2}$. Therefore, we prove the exact constants in $(Y \setminus M(s.In \setminus$

$s.DIn))$ can be re-computed as objects in facts by using the expression of $\Delta_s$ defined in Table 1.

$$Y \setminus M(s.In \setminus s.N) = \quad (14)$$
$$H(j.In + j.D + [j.idx]) \setminus M(s.In \setminus s.N) = \quad (15)$$
$$M(j.In + j.D + [j.idx]) \setminus M(s.In \setminus s.N) = \quad (16)$$
$$M((j.In + j.D + [j.idx]) \setminus s.In) +$$
$$(M(s.N) \cap M(s.In) \cap Y) \subseteq \quad (17)$$
$$M(s.D) + M(s.N \setminus s.D). \quad (18)$$

The expressions above prove the ground truth (expression 14) is included in the definition of $\Delta_s$ shown in expression 18. The $DIn$ and $NIn$ are shorted as $D$ and $N$ respectively for simplicity. From expression 14 to 15, the $Y$ is expanded as defined above. Lemma 7 shows how expression 15 can be transferred to 16, while expression 17 is computed by some set operators. The transformation from 17 to 18 is because $(j.In + j.D + [j.idx]) \setminus s.In$ is exactly what will be passed to $s.D$ if $s$ is *stable*; and for stable $s$, $s.In \cap s.D = \emptyset$, $s.N \subseteq s.In + s.D$. Additionally, since only $(j.In + j.D + [j.idx]) \setminus s.In$ is added to $SCC_X$ for $s \in SCC_X$ and $s.St = stable$, the correctness is also very intuitive.

For $s \in SCC_X$ and $s.St = new$, $\bigcup_{n \in F(s)} \{M(n) \times M(n.In \setminus s.N)\} \subseteq I^{T_1}$, and should be eliminated from $\Delta^{T_2}$. Therefore, the expected facts in $\Delta^{T_2}$ are: $\bigcup_{n \in F(s)} \{M(n) \times (Y \setminus M(n.In \setminus s.N))\}$, that is included in $\Delta^{T_2}$ as proved below.

$$Y \setminus M(n.i \setminus s.N) = \quad (19)$$
$$H(j.In + j.D + [j.idx]) \setminus M(n.In \setminus s.N) = \quad (20)$$
$$M(j.In + j.D + [j.idx]) \setminus M(n.In \setminus s.N) = \quad (21)$$
$$M((j.In + j.D + [j.idx]) \setminus n.In) +$$
$$(M(s.N) \cap M(n.In) \cap Y) \subseteq \quad (22)$$
$$M(s.D \setminus n.In) + M(s.N \setminus (s.D \setminus n.In)). \quad (23)$$

This process is similar to the process from expression 14 to 18. There is only a slight difference from expression 22 to 23: for a *new* SCC $s$, $j.In + j.D + [j.idx]$ is added to $s.D$ directly without minus $s.In$, so $j.In + j.D + [j.idx] \subseteq s.D$. And $s.N \subseteq s.D + s.In$.

Therefore, $\Delta^{T_2}$ can generate additional facts as we expected. Since the insertion is monotonous, we only add new $DIn$ to $SCC_X$, and do not delete anything. Therefore, facts in $\Delta^{T_1}$ is still included in $\Delta^{T_2}$.

(3) $I^{T_2} \cup \Delta^{T_2} = r^\infty(N \cup \Delta_2)$: the proof of $I^{T_2} \cup \Delta^{T_2}$ is intuitive. Use $Y$ and $SCC_X$ defined above, we expect $Y$ is included in the objects of every $s \in SCC_X$. For a *stable* $s \in SCC_X$, $j.In + j.D + [j.idx] \subseteq s.In + s.D$. Refer to table 1, the objects of facts in $I^{T_2} \cup \Delta^{T_2}$ are computed by the operator $H$, therefore $Y = H(j.In + j.D + [j.idx]) \subseteq H(s.In + s.DIn)$. For a *new* $s \in SCC_X$, $j.In + j.D + [j.idx] \subseteq s.D$, so therefore $Y = H(j.In + j.D + [j.idx]) \subseteq H(s.DIn)$.

- **Case 2: [non-tree] back edge**:

(1) $I^{T_2} = I$: The pre-condition is that $I^{T_1}$ is correct. Let $s$ be the *new* SCC after merging, for $n \in F(s)$, $I_n = \{M(n) \times M(n.In \setminus n.N)\} \subseteq I^{T_1}$ is the facts generated for $n$ before merging. In $I^{T_2}$, $I'_n = \{M(n) \times M(n.In \setminus s.N)\}$. We prove $I_n = I'_n$ below.

For every $n_1 \neq n_2$ and $n_1, n_2 \in F(s)$, $n_2.N \cap n_1.In \subseteq n_1.N \cap n_1.In$. This could be easily proved by contradiction. Assume there exists $n_2$ s.t. there exists an interval $In \subseteq (n_2.N \cap n_1.In)$ and $In \not\subseteq (n_1.N \cap n_1.In)$. Since $In \subseteq n_2.N$, there exists at least one new node introduced in the interval $In$. $In \subseteq n_1.In$ implies that $n_1$ could reach this new node, if the previous steps are correct, then $In \subseteq n_1.N$, so $In \subseteq n_1.N \cap n_1.In$, which contradicts our assumption.

In line 15 of algorithm 4, $s.N = \bigcup_{n \in F(s)} \{n.N\}$. Therefore, for every node $n \in F(s)$, $s.N \cap n.In = n.N \cap n.In$. Thus $I_n = I'_n$ and $I^{T_2}$ stays unchanged.

(2) $\Delta^{T_2} = r^\infty(N \cup \Delta_2) \setminus I$: There are two parts of the facts we expect from $\Delta^{T_2}$, in addition to $\Delta^{T_1}$. (a) for $n \in F(s)$, $n$ now can reach all the resources contained in $F(s)$, as well as all the descendants of $F(s)$ in $I^{T_1} \cup \Delta^{T_1}$. Formally, the resources $n$ can reach is $H(s.D)$ in which $s.D$ is defined in line 14 of algorithm 4. If $n.St = stable$ before merging, in $I^{T_1}$, $n$ can reach $M(n.In \setminus s.N)$, which should not be included in $\Delta^{T_2}$. Formally, for $n \in F(s)$, $n.St = stable$, we prove the expected constants described above can be re-computed by using the expressions in 1.

$$H(s.D) \setminus M(n.In \setminus n.N) = \quad (24)$$
$$H(s.D) \setminus M(n.In \setminus s.N) = \quad (25)$$
$$M(s.D) \setminus M(n.In \setminus s.N) = \quad (26)$$
$$M(s.D \setminus n.In) + M(s.N \cap s.D \cap n.In) = \quad (27)$$
$$M(s.D \setminus n.In) + M(s.N \setminus (s.D \setminus n.In)) \quad (28)$$

The same can be stated for $n \in F(s), n.St = new$ or *dropped* before merging, the only difference is expression 24 will be $H(s.D) \setminus M(n.In \setminus s'.N)$, in which $s'$ is the original representative node s.t. $n \in F(s')$. The $s'.N$ will be added to $s.N$ and $n.In$ stay unchanged, so the transformation from expression 24 to 25 still stands. The transformation from expression 25 to expression 26 is because $s.D$ is the union of intervals that satisfies the conditions stated in lemma 5 and lemma 7. Therefore $H(s.D)$ is safe to convert to $M(s.D)$. As stated above, for every $n \in F(s)$, correct facts can be generated in $\Delta^{T_2}$.

(b) The other part of the expected facts in $\Delta^{T_2}$ is that: if $a$ is an ancestor of $s$, then $a$ can reach constants in $H(s.D)$, too. This is realised by passing $s.D$ to every ancestor of $s$. For each ancestor $a$, $H(s.D)$ minus the resources $a$ could reach in $I^{T_1}$ is expected in $\Delta_a$. The rest of the proof is similar to the case of cross edge and is not repeated here.

(3) $I^{T_2} \cup \Delta^{T_2} = r^\infty(N \cup \Delta_2)$: refer to table 1, $H(s.D)$ is directly included in the expression of $I^{T_2} \cup \Delta^{T_2}$, so the correctness is intuitive.

- **Case 3: [tree-edge] new child**: In this case, we maintain the data structure in the TC storage scheme by inserting the new node $j$ to $L$, and choosing an empty slot from the tree interval of $i$ to initialise $j$'s interval. And we pass

the interval of $j.TIn$ to $i.NIn$, as well as the ancestors of $i$. Only the SCC $j$ is included in $L_{j.TIn}$.

(1) $I^{T_2}= I$: for $j \in L$, nothing will generate since $j.In$ is empty; For $s \in L$ and $s \neq j$, $s.N$ is excluded so $j$ is excluded. Therefore, $I^{T_2}$ will generate the same facts as $\Delta^{T_1}$.

(2) $\Delta^{T_2}= r^\infty(N \cup \Delta_2)\backslash I$: In addition to $\Delta^{T_1}$, new facts should be included in $\Delta^{T_2}$ are: $\{X \times j'\}$, in which $X$ includes constants in $i$, as well as all the ancestors of $i$ in $G$. Formally, $X = \{x \mid (x, i') \in I^{T_1} \cup \Delta^{T_1}\}$.

Define $SCC_X = \{s \mid s \in L, H(s) \subseteq X\}$. The definition of $SCC_X$ guarantees $i.N$ will be passed to every $s \in SCC_X$ as a $s.N$ by the *propagate* function. Refer to Table 1, $j \in L$ will generate nothing since $L_{j.D}$ is empty. For every $s \in L$, but $s \notin SCC_X$, the stored intervals stay unchanged so $\Delta_s$ is the same as that in $\Delta^{T_1}$. For $s \in L$ and $s \in SCC_X$, $j.TIn = j.DIn + [j.idx] \subseteq s.N$, so $j' \in M(s.N)$. If $s$ is *stable*, since $s.N \subseteq s.In + s.D$, $j$ can be generated as an object in the fact, by using the expression defined in Table 1.

If $s$ is *new*, according to lemma 4, for every $n \in F(s)$, $s.N \subseteq n.In + s.D$, so $j$ can be generated as an object of the facts, by using the expression shown in Table 1; for subjects, $H(s) = \bigcup_{n \in F(s)} M(n)$. Therefore, $\Delta^{T_2}$ can generate $\{X \times j'\}$ accurately in addition to $\Delta^{T_1}$.

(3) $I^{T_2} \cup \Delta^{T_2}= r^\infty(N \cup \Delta_2)$: For *stable* $s \in SCC_X$, $s.N \subseteq s.In + s.D$. So $j' \in M(s.N) = H(s.N) = H(s.In + s.D)$. For *new* $s \in SCC_X$, $s.N \subseteq s.In + s.DIn = s.DIn$, so $j' \in M(s.N) = H(s.N) = H(s.DIn)$. Therefore, expressions in Table 1 can include $(X, j')$ in addition to $I^{T_1} \cup \Delta^{T_1}$. Also, since $j$ is the only $SCC$ that in $L_{s.N}$, the correctness can also be easily proved.

- **Case 4: [tree-edge] new parent**: In this case, a new parent $i'$ is introduced for $j'$, which will cause $i'$ can reach everything that $j'$ can reach, as well as $j'$ itself. Formally, facts $\{i' \times X\}$ in which $X = \{x \mid (j', x) \in I^{T_1} \cup \Delta^{T_1}\} \cup \{j'\}$, should be included in $\Delta^{T_2}$.

(1) $I^{T_2}= I$: we only changed the intervals of $i$; and inserted $i$ to L. Since $i.St = stable$, according to expressions in Table 1, $\{(M(i) = i') \times (i.In \backslash i.N) = \emptyset\}$ will be generated. So for $i$, there are no facts generated. For other SCC in $L$, the generated facts are the same as $I^{T_1}$. Therefore, nothing will be generated in addition to $I^{T_1}$.

(2) $\Delta^{T_2}= r^\infty(N \cup \Delta_2)\backslash I$: the $In$, $DIn$ and $NIn$ stay unchanged for all the $s \in L$ except $i$. For these nodes, they can already generate facts correctly.

The additional facts we expect from $\Delta^{T_2}$ is $(i' \times X)$. $X$ includes the nodes that $j'$ can reach in $I^{T_1} \cup \Delta^{T_1}$ and $j'$, i.e., $H(j.In + j.D + [j.idx])$. Since $i.D = j.In + j.D + [j.idx]$, $\Delta_i$ will output $\{i' \times X' \mid X' = M(i.D)\}$. We prove $X = X'$ to show the correctness of $I^{T_2} \cup \Delta^{T_2}$, i.e., $H(j.i + j.D + [j.idx]) = M(j.i + j.D + [j.idx])$, which can be proved by lemma 7. Therefore, $\Delta^{T_2}$ can generate facts correctly.

(3) $I^{T_2} \cup \Delta^{T_2}= r^\infty(N \cup \Delta_2)$: the intervals stored for all the other SCC except $i$ are not changed. The additional facts we expect is $\{i' \times X\}$. According to expressions in Table 1, $\{i' \times H(i.D)\}$ can precisely output $(i' \times X)$ defined above.

- **Case 5: [tree-edge] new child and parent**: this case can be proved similarly to case 3, so the proof is omitted here.
- **Case 6: [non-tree] forward edge**:

(1) $I^{T_2}= I$: nothing is changed.

(2) $\Delta^{T_2}= r^\infty(N \cup \Delta_2)\backslash I$: since $(i', j') \in I^{T_1} \cup \Delta^{T_1}$, the insertion of this edge will not influence the closure. So $\Delta^{T_1}= r^\infty(N \cup \Delta_1)\backslash I = r^\infty(N \cup \Delta_2)\backslash I = \Delta^{T_2}$.

(3) $I^{T_2} \cup \Delta^{T_2}= r^\infty(N \cup \Delta_2)$: same as (2) above.

Combining all the cases above, *TC_add* can maintain the data structure correctly for a single insertion. □

Then we show that the function execution of *TC_add*($\Delta$) with the index 3 in Table 2 can update the data structure correctly so that $I^T$ and $\Delta^T$ can be accessed correctly.

**Claim 3** *The TC_add($\Delta$) function (with the index 3 in Table 2) can maintain the data structure correctly so that $I^T$ and $\Delta^T$ can return the facts defined in Table 2.*

**Proof.** We prove this claim by induction.

1. *Induction base*: Let the edges that have been processed be $\Delta'$. Before any edge $e \in \Delta$ is inserted, i.e., $\Delta' = \emptyset$, the data structure in TC storage scheme remains the same as after the last *merge* function. Therefore, $I^T= r^\infty(N)$ as before, $\Delta^T= \emptyset = r^\infty(N) \setminus r^\infty(N) = r^\infty(N \cup \Delta')\backslash r^\infty(N)$. $I^T \cup \Delta^T = r^\infty(N) = r^\infty(N \cup \Delta')$. Therefore, before any edge is inserted, $I^T$ and $\Delta^T$ can access facts as expected.

2. *Induction step*: If the previous insertions of edges in $\Delta'$ are correct, the pre-condition of lemma 8 is satisfied. Also, the correctness of inserting a single edge $e$ is proved by lemma 8. Therefore, we have the insertion of $\Delta' = \Delta' + \{e\}$ is correct.

As stated above, *TC_add*($\Delta$) can insert $\Delta$ correctly so that $I^T$ and $\Delta^T$ can return the facts as expected. □

**Claim 4** *The merge function (with the index 4 in Table 2) can maintain the data structure correctly so that $I^T= I^T \cup \Delta^T = r^\infty(N \cup \Delta)$, $\Delta^T$ returns nothing.*

**Proof.** Claim 3 shows *TC_add*($\Delta$) can maintain the data structure correctly. Let $s$ and $s'$ be the same SCC before and after calling *merge*. We will prove this claim by showing $I_s \cup \Delta_s = I_{s'}$. If $s.St = stable$, $I_s \cup \Delta_s = \{M(s) \times H(s.In + s.DIn)\}$. Then $s'.St = stable$, and $I_{s'} = \{M(s') \times M(s'.In \backslash s'.N)\}$. In the *merge* function, the constants contained is not changed for a stable SCC, so $M(s) = M(s')$. For the intervals of $s$ and $s'$, we have: $s'.In = s.In + s.D$, and $s'.N = s'.D = \emptyset$. Therefore, $s'.In \backslash s'.N = s'.In = s.In + s.D$. The operator $H$ will collect $M(n)$ for $n \in s.In + s.D$ and $n.St = stable$, and $nM(n)$ if $n.St = new$. The *dropped* SCCs in this interval are ignored. In the *merge* function, all the *dropped* SCCs are removed from $L$, and for *new* SCC $n \in L$, let $n'$ be the same node after the call, $M(n') = nM(n) = H(n)$. For *stable* SCC $n \in L$, $M(n') = M(n) = H(n)$. Therefore,

$H(s.In + s.D) = M(s'.In \setminus s'.N)$. As stated above, for the SCC $s$ that was *stable* before the call, $I_s \cup \Delta_s = I_{s'}$.

On the other hand, if $s.St = new$, then $s'.St = stable$. For $I_s \cup \Delta_s$, we have $\{nM(s) \times H(s.D)\}$. For $I_{s'}$, we have $\{M(s') \times s'.In \setminus s'.N\}$. In the *merge* function, the $M(s')$ is updated to $nM(s)$. And $s'.In \setminus s'.N = s'.In = s.In + s.D = s.D$. Similar to above, $H(s.D) = M(s'.In \setminus s'.N)$. Therefore, for SCC $s$ that was *new*, $I_s \cup \Delta_s = I_{s'}$. In conclusion, let $T_1(\text{'label'})$ be the facts returned before calling the *merge* function with index 4 but after calling *TC_add* with index 3, and $T_2(\text{'label'})$ be the facts returned after calling *merge*, we have $I^{T_1} \cup \Delta^{T_1} = I^{T_2}$.

In addition, since the $NIn$ and $DIn$ of all SCC $s \in L$ have been emptied, $\Delta^T = \emptyset$, and $I^T \cup \Delta^T = I^T$, as stated in the claim 4. $\square$

By showing the correctness of the first few calls of the TC storage scheme in claim 1 $\sim$ 4, we show that the functions of the TC storage scheme can be used correctly in the multi-scheme reasoning algorithm presented in algorithm 2 of the main submission. Define $O_T$ as the union of the facts that have been inserted into $T$ before the last call of *merge* in the call sequence of the scheme $T$, and define $N_T$ as the facts that have been inserted into $T$ after the last call of *merge*. The set of facts in different domains 'I', '$\Delta$' and '$I \cup \Delta$' is defined as follows:

$$I^T = \Pi_T^\infty(O_T) \tag{29}$$

$$\Delta^T = \Pi_T^\infty(O_T \cup N_T) \setminus I^T \tag{30}$$

$$I^T \cup \Delta^T = \Pi_T^\infty(O_T \cup N_T) \tag{31}$$

The $N_T$ defined above is $C_T$ that is populated in the *schedule* function, and $\Delta_n^T = C_T = N_T \subseteq \Pi_T^\infty(O_T \cup N_T) \setminus I^T$. Therefore, we have the following lemma:

**Lemma 9** *The derive function of the TC scheme satisfies the property stated in the main submission by updating $\Delta^T$ as:*

$$\Delta^T = \Delta_n^T \cup \Pi_T^\infty[I \cup C_T] \setminus I \tag{32}$$

**Lemma 10** *The merge function of the TC scheme updates $I^T$ and $\Delta^T$ as defined in the main submission.*

## Union Storage Scheme

The complete version of functions for union schemes is presented in algorithm 7. For a union storage scheme $T$, the order of $T$'s functions being called follows the same pattern as the TC storage scheme (refer to the section of the TC storage scheme above). The union storage scheme is literally the "union" of facts with different supporting predicates. For example, $I^T$ in algorithm 7 is defined as the union of facts in $I_p$, with the predicate translating to the union predicate $U$, in which $p \in P_T \cup SP_T$. Therefore, $\Delta^T$ is correct if the supporting predicates in $P_T \cup SP_T$ are correctly maintained. The $\Delta^T$ can also be represented as the union of facts in $\Delta_p$, or equivalently, the facts that are not in $I_p$ for every $p \in P_T \cup SP_T$. In algorithm 7, the *schedule* will filter the facts that satisfy this condition (line 10). Then, the facts with the predicate $U$ are added to $\Delta_n^T$ if they cannot be recovered by other facts (line 11); The facts with the predicate in $SP_T$ are added to $C_T$ (line 12). The *derive* function changes the

---

Algorithm 7: Functions of Union Table

1: $T$: a union table;    $t$: a fact;    $L_T$: the fact list in $T$.
2: $D_T$: implemented as $L_T$.    Assume $P_T = \{U\}$.
3: $\Pi_T = \{r \mid r = p(\vec{x}) \rightarrow U(\vec{x}) \in \Pi\}$.
4: **procedure** COMPUTE $I^T$
5:    $I^T := \{t \in L_T \mid \text{the label of } t = \text{'}I\text{'}\}$
6:    **for** $p \in SP_T$, **for** $t \in I_p$ **do**
7:      $I^T := I^T \cup \{U(\vec{x}).sub(t)\}$
8: **procedure** $T.schedule(t)$
9:    **if** $t.p \notin P_T \cup SP_T$ **then** return
10:    **if** $p(\vec{x}).sub(t) \notin I_p$ for any $p \in SP_T \cup P_T$ **then**
11:      **if** $t.p \in P_T$ **then** add $t$ to $L_T$ as '$\Delta_n$'
12:      **if** $t.p \in SP_T$ **then** $C_T := C_T \cup \{t\}$
13: **procedure** $T.merge()$
14:    mark facts in $L_T$ with '$\Delta$' as '$I$'
15: **procedure** $T.derive()$
16:    mark facts in $L_T$ with '$\Delta_n$' as '$\Delta$'
17:    $\Delta^T := \{t \in L_T \mid \text{the label of } t = \text{'}\Delta\text{'}\}$
18:    $\Delta^T := \Delta^T \cup \{U(\vec{x}).sub(t) \mid t \in C_T\}$,    $C_T = \emptyset$

---

facts with '$\Delta_n$' to '$\Delta$', so $\Delta_n^T$ is included in $\Delta^T$. The application of rules in $\Pi_T$ is realised by translating the predicate to $U$ (line 18). Therefore, we can have the following lemma:

**Lemma 11** *The derive function of the union scheme updates $\Delta^T$ as $\Delta^T = \Delta_n^T \cup \Pi_T[I, C_T] \setminus I$.*

The *merge* function will change the facts with the label '$\Delta$' to '$I$'. For facts in $\Delta^T \setminus L_T$ that are not managed by $L_T$ but recovered from other facts, the *merge* action is realised by calling the *merge* function of corresponding schemes, which is done in line 8 of the multi-scheme reasoning algorithm. Therefore, we have the following lemma:

**Lemma 12** *The merge function of the union scheme updates $I^T$ to $I^T \cup \Delta^T$, empties $\Delta^T$.*

## Correctness of Multi-Scheme Reasoning

In this section, we show the correctness of the multi-scheme reasoning algorithm by showing algorithm 2 of the main submission is equivalent to the modular seminaïve algorithm. In line 4 of algorithm 2, the *derive* function of schemes computes the $\Delta$ just like the $A$ in the modular seminaïve algorithm. Differently, the $\Delta$ in the multi-scheme algorithm does not strictly eliminate the input facts that are used to derive $\Delta$. In the multi-scheme algorithm, the input facts are stored in set $\Delta_n^T$, and are included in $\Delta$. But in the modular algorithm, the input facts are $\Delta$, and the derivations are $A$, which is strictly distinguished from $\Delta$. Therefore, in the multi-scheme algorithm, when scheduling the $\Delta$ to different schemes in line 7, the original input facts that are used to derive $\Delta$ will be scheduled again. But this will only influence the effectiveness, not the correctness. The multi-scheme reasoning algorithm still considers the same rule instances as the modular approach, just with some repetitions. For the sake of effectiveness, the *schedule* function of schemes is designed to try to eliminate the repetitions when it is possible. For example, for a TC scheme $T$, if a fact $t$

|  | $Q_0$ (140M) | $Q_1$ (140M) | $Q_2$ (643) | $Q_3$ (428) | $Q_4$ (325) | $Q_5$ (5) | $Q_6$ (140M) | $Q_7$ (752k) | $Q_8$ (1M) | $Q_9$ (12) | $Q_{10}$ (7) |
|---|---|---|---|---|---|---|---|---|---|---|---|
| Standard | 8.38 | 27.70 | 0.03 | 0.03 | 0.03 | 0.03 | 15.00 | 0.10 | 0.13 | 0.03 | 0.03 |
| MultiTable | 22.43 | 20.64 | 0.03 | 0.03 | 0.03 | 0.03 | 42.15 | 0.24 | 0.35 | 0.03 | 0.03 |
|  | $Q_0$ (280M) | $Q_1$ (280M) | $Q_2$ (753k) | $Q_3$ (1M) | $Q_4$ (337) | $Q_5$ (12) | $Q_6$ (280M) | $Q_7$ (753k) | $Q_8$ (1M) | $Q_9$ (337) | $Q_{10}$ (12) |
| Standard | 25.50 | 77.00 | 0.28 | 0.42 | 0.03 | 0.03 | 36.00 | 0.10 | 0.15 | 0.03 | 0.03 |
| MultiTable | 338.90 | 393.00 | 0.61 | 0.89 | 0.03 | 0.03 | 416.23 | 1.97 | 2.93 | 0.03 | 0.03 |

Table 3: Running time of Query Answering. The upper 11 queries are about the transitive predicate; while the lower 11 queries are about the union predicate.

with the predicate in $P_T$ is scheduled, $T$ will check whether the fact $t$ is in $I^T \cup \Delta^T$. If $t \in I^T$, then $t$ is already considered in the previous round. If $t \in \Delta^T$, then $t$ is just computed in line 4 in the current round. The TC scheme reaches the closure directly, so there is no need to schedule $t$ again. For the plain table, facts in it are assigned a unique index to identify each distinctive fact. So facts managed by the plain table are easily distinguished to avoid repetition when they are scheduled for schemes. But for facts stored in the TC and union scheme, they are not intuitively distinguished from the input facts in $\Delta_n^T$. However, transitive rules and union rules usually extend input facts to a large extent, and the inefficiency caused is limited. Based on the discussion above, we have the following lemma:

**Lemma 13** *A fact can be derived and represented in relevant schemes by the multi-scheme algorithm if and only if it can be derived by the modular seminaïve algorithm.*

## Evaluations of QA

The complete results of query answering experiments are shown in Table 3. The queries used for testing are shown in Table 4 and Table 5 in which following prefixes are used:

1. skos:<http://www.w3.org/2004/02/skos/core#>,
2. db:<http://dbpedia.org/resource/>,
3. term: <http://purl.org/dc/terms/ >.

## The Availability of Benchmarks

As described in the main text, we utilise three benchmarks for evaluation: DAG-R, Relations, and DBpedia, which were originally used by Hu, Motik, and Horrocks (2022). For DAG-R and Relations, we use the same programs and datasets that were made publicly available by the authors here[2]. For DBpedia, we access the program from the same website, but use a more complete version of the DBpedia facts whereas the website only provides a subset. All the benchmarks mentioned above are available online[3].

---
[2]https://krr-nas.cs.ox.ac.uk/2021/modular-reasoning/
[3]https://xinyuezhang.xyz/TCReasoning/

| | Query |
|---|---|
| TC $Q_0$ | Select (count(*) as ?cnt) where { <br>   ?x skos:narrowerTransitive ?y . <br> } |
| TC $Q_1$ | Select (count(*) as ?cnt) where { <br>   ?x a skos:Concept . <br>   ?x skos:narrowerTransitive ?y . <br> } |
| TC $Q_2$ | Select (count(*) as ?cnt) where { <br>   ?x term:subject db:Reproducibility . <br>   ?x skos:narrowerTransitive ?y . <br> } |
| TC $Q_3$ | Select (count(*) as ?cnt) where { <br>   ?x term:subject db:Uncertainty . <br>   ?x skos:narrowerTransitive ?y . <br> } |
| TC $Q_4$ | Select (count(*) as ?cnt) where { <br>   ?x term:subject <br>     db:Supercritical_angle_fluorescence_microscopy . <br>   ?x skos:narrowerTransitive ?y . <br> } |
| TC $Q_5$ | Select (count(*) as ?cnt) where { <br>   ?x term:subject db:Micrograph . <br>   ?x skos:narrowerTransitive ?y . <br> } |
| TC $Q_6$ | Select (count(*) as ?cnt) where { <br>   ?y a skos:Concept . <br>   ?x skos:narrowerTransitive ?y . <br> } |
| TC $Q_7$ | Select (count(*) as ?cnt) where { <br>   ?y term:subject db:Reproducibility . <br>   ?x skos:narrowerTransitive ?y . <br> } |
| TC $Q_8$ | Select (count(*) as ?cnt) where { <br>   ?y term:subject db:Uncertainty . <br>   ?x skos:narrowerTransitive ?y . <br> } |
| TC $Q_9$ | Select (count(*) as ?cnt) where { <br>   ?y term:subject <br>     db:Supercritical_angle_fluorescence_microscopy . <br>   ?x skos:narrowerTransitive ?y . <br> } |
| TC $Q_{10}$ | Select (count(*) as ?cnt) where { <br>   ?y term:subject db:Micrograph . <br>   ?x skos:narrowerTransitive ?y . <br> } |

Table 4: The queries with the TC predicate.

| | Query |
|---|---|
| U $Q_0$ | Select (count(*) as ?cnt) where { <br>   ?x skos:semanticRelation ?y . <br> } |
| U $Q_1$ | Select (count(*) as ?cnt) where { <br>   ?x a skos:Concept . <br>   ?x skos:semanticRelation ?y . <br> } |
| U $Q_2$ | Select (count(*) as ?cnt) where { <br>   ?x term:subject db:Reproducibility . <br>   ?x skos:semanticRelation ?y . <br> } |
| U $Q_3$ | Select (count(*) as ?cnt) where { <br>   ?x term:subject db:Uncertainty . <br>   ?x skos:semanticRelation ?y . <br> } |
| U $Q_4$ | Select (count(*) as ?cnt) where { <br>   ?x term:subject <br>     db:Supercritical_angle_fluorescence_microscopy . <br>   ?x skos:semanticRelation ?y . <br> } |
| U $Q_5$ | Select (count(*) as ?cnt) where { <br>   ?x term:subject db:Micrograph . <br>   ?x skos:semanticRelation ?y . <br> } |
| U $Q_6$ | Select (count(*) as ?cnt) where { <br>   ?y a skos:Concept . <br>   ?x skos:semanticRelation ?y . <br> } |
| U $Q_7$ | Select (count(*) as ?cnt) where { <br>   ?y term:subject db:Reproducibility . <br>   ?x skos:semanticRelation ?y . <br> } |
| U $Q_8$ | Select (count(*) as ?cnt) where { <br>   ?y term:subject db:Uncertainty . <br>   ?x skos:semanticRelation ?y . <br> } |
| U $Q_9$ | Select (count(*) as ?cnt) where { <br>   ?y term:subject <br>     db:Supercritical_angle_fluorescence_microscopy . <br>   ?x skos:semanticRelation ?y . <br> } |
| U $Q_{10}$ | Select (count(*) as ?cnt) where { <br>   ?y term:subject db:Micrograph . <br>   ?x skos:semanticRelation ?y . <br> } |

Table 5: The queries with the Union predicate.



\end{document}